\documentclass[prb, floatfix, nobibnotes, reprint, superscriptaddress]{revtex4-1}

\usepackage[version=4]{mhchem} 

\usepackage[T1]{fontenc}
\usepackage{textgreek}  
\usepackage{color}
\usepackage{graphicx}
\usepackage{dcolumn}
\usepackage{bm}
\usepackage[table,xcdraw]{xcolor}

\usepackage{amsfonts} 
\usepackage{amsmath}
\usepackage{footnote}
\usepackage{mathtools}

\usepackage{booktabs}
\usepackage{array,multirow}
\usepackage{ragged2e}

\begin{document}
\title{\papertitle}
\title{Hammett-Inspired Product Baseline for Data-efficient $\Delta$-ML in Chemical Space}

\author{V. Diana Rakotonirina}
\affiliation{Department of Materials Science and Engineering, University of Toronto, St. George campus, Toronto, ON, Canada}
\author{Marco Bragato}
\affiliation{Faculty of Physics, University of Vienna, Kolingasse 1416, AT1090 Wien, Austria}
\author{Guido Falk von Rudorff}
\affiliation{Department of Chemistry, University Kassel,Heinrich-Plett-Str.40, 34132 Kassel, Germany \\
Center for Interdisciplinary Nanostructure Science and Technology (CINSaT)}
\author{O. Anatole von Lilienfeld}
\email{anatole.vonlilienfeld@utoronto.ca}
\affiliation{Machine Learning Group, Technische Universität Berlin}
\affiliation{Vector Institute for Artificial Intelligence, Toronto, ON, M5S 1M1, Canada}
\affiliation{Department of Chemistry, University of Toronto, St. George campus, Toronto, ON, Canada}
\affiliation{Berlin Institute for the Foundations of Learning and Data, Berlin, Germany}
\affiliation{Acceleration Consortium, University of Toronto. 80 St George St, Toronto, ON M5S 3H6}
\affiliation{Department of Materials Science and Engineering, University of Toronto, St. George campus, Toronto, ON, Canada}
\affiliation{Department of Physics, University of Toronto, St. George campus, Toronto, ON, Canada}





\begin{abstract}
Data-hungry machine learning methods have become a new standard to efficiently navigate chemical compound space for molecular and materials design and discovery. 
Due to the severe scarcity and cost of high-quality experimental or synthetic simulated training data,
however, data-acquisition costs can be considerable. 
Relying on reasonably accurate approximate legacy baseline labels with low computational complexity 
represents one of the most effective strategies to curb data-needs, 
e.g.~through $\Delta$-, transfer-, or multi-fidelity learning. 
A surprisingly effective and data-efficient baseline model is presented
in the form of a generic coarse-graining Hammett-Inspired Product (HIP) {\em Ansatz},  
generalizing the empirical Hammett equation towards arbitrary systems and properties. 
Numerical evidence for the applicability of HIP includes  solvation free energies of molecules, 
formation energies of quaternary elpasolite crystals, 
carbon adsorption energies  on heterogeneous catalytic surfaces, 
HOMO-LUMO gaps of metallorganic complexes, 
activation energies for S$_\text{N}$2 reactions, and catalyst-substrate binding energies in cross-coupling reactions.
After calibration on the same training sets, 
  HIP yields an effective baseline for improved $\Delta$-machine learning 
 models with superior data-efficiency when compared to previously introduced 
 specialised domain-specific models.
\end{abstract}

\maketitle

\section{Introduction}
\begin{figure}[h!]
	    \centering
    \includegraphics[width=0.45\textwidth]{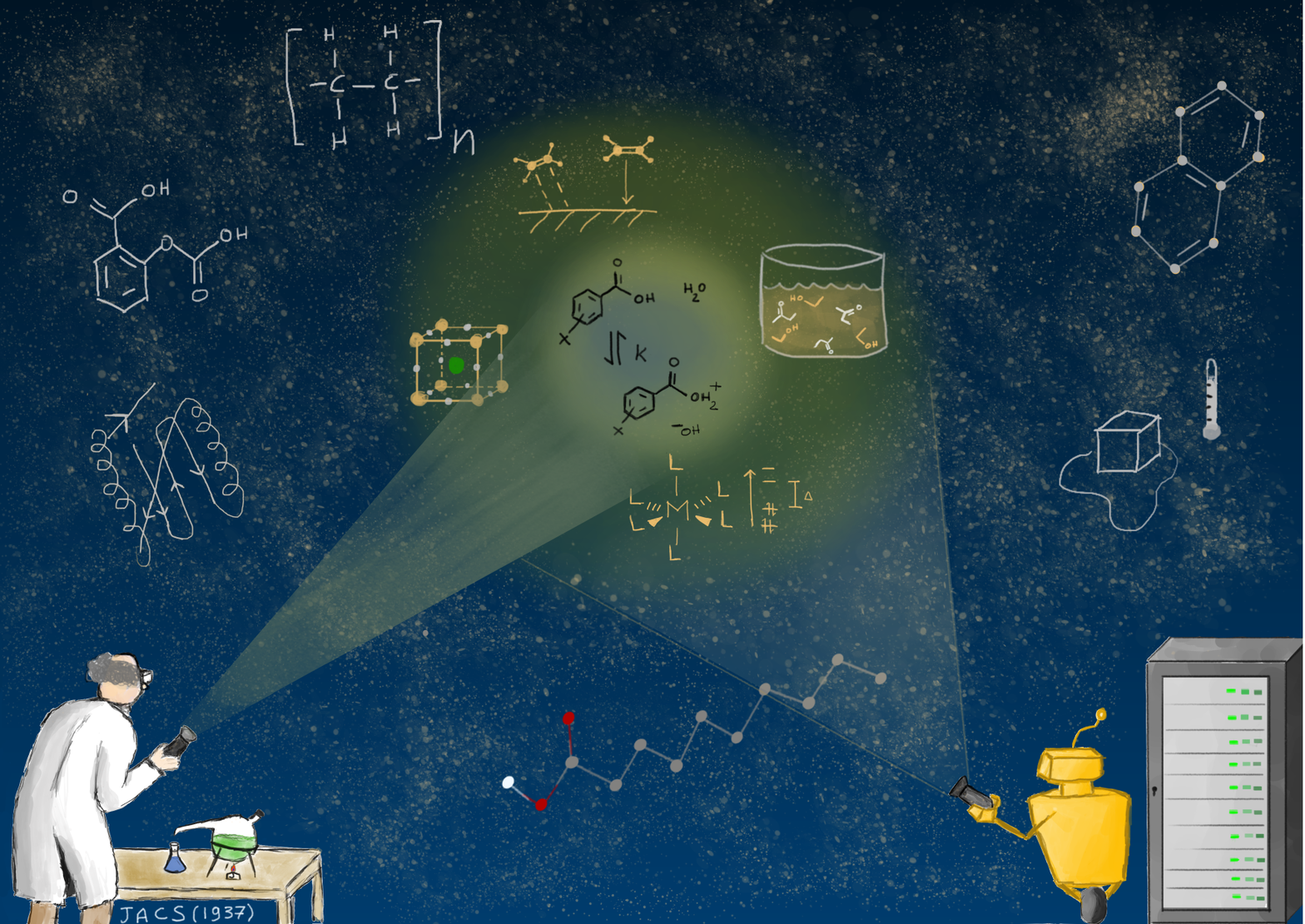}
    \caption{Generalizing Hammett's equation (Left) towards our coarse-grained product Ansatz for arbitrary chemical spaces and properties (Right).}
    \label{fig:TOC}
\end{figure}

In the late '30s, Luis Hammett proposed a simple yet effective linear free energy relationship for evaluating substituent effects on different reactions: a simple product between two scalar quantities\cite{Hammett1935,Hammett1937}.
Product ansatz are not a new concept in chemistry, they have been around since the dawn of quantum chemistry; the Hartree product, for example, played an important role in the early evaluation of orbital energies. 
Similarly, the idea of relying on simple scalars to describe physically and chemically interesting properties is not new either. 
Volcano plots are one of such examples: they were independently developed in the 50s by Gerishers\cite{Gerischer1958} and Parson\cite{Parsons1958} as a way to assess the quality of heterogeneous catalysts based on the adsorption strength, in accordance with the Sabatier principle, and, interestingly, only in more recent years have been treated theoretically\cite{Norskov2004,Bligaard2004,Norskov2005}.

The Hammett equation, despite being studied and expanded during the last century, in order to encompass resonance\cite{Swain1968}, nucleophilicity\cite{Swain1953}, steric\cite{Taft1952, Taft1952-2, Taft1953,Santiago2016} and solvent effects\cite{Jahagirdar1988, Kondo1969, Grunwald1948, Winstein1951,Dobrowolski2018,Song2006}, has been used almost exclusively in the field of reactions and substituent effects, mainly on benzene-like molecules, with the exception of studies on the correlation between Hammett parameters and other experimentally observed quantities\cite{Masui1993,Buszta2019,Kumar2018,Venkataraman2007}.
However, such restriction is not imposed at any point by the Hammett equation and in a previous work\cite{Bragato2020} we showed how the model can be enhanced to be more robust, interpretable and general enough to encompass a wider area of the Chemical Compound Space (CCS).

In the last decades, the exploration of the CCS saw the rise of statistical methods, mostly machine learning (ML), with established methods like Artificial Neural Networks (ANN) and Kernel Ridge Regression (KRR). 
More recently, other physics-based approaches, like Alchemical Derivation, focused on exploring a smaller portion of the CSS starting from well-defined initial points\cite{vonRudorff2019,vonRudorff2020,vonRudorff2021}.
The core effort of all of the above methods has been put into pushing for ever greater accuracy, often at the cost of transferability of the final model on different systems or properties.
While these approaches have proved time and time again to be successful in being very accurate, they often suffer from transferability issues and might require a lot of adjustment when the scope of the exploration changes. 

A simpler model like the Hammett equation can be a useful complement to the aforementioned ones since what it lacks in accuracy makes up for in transferability, much like fragment-based descriptors have been used to encode transferable patterns and mitigate combinatorial complexity in inorganic materials\cite{isayev2017universal}.
This structured simplicity allows such models to extend predictive capabilities with minimal data, especially in early-stage screening or low-data regimes.
The main goal of the Hammett model, therefore, is not to reach extremely accurate predictions, but rather to easily helm an initial exploration of the CCS around the desired property.
To recover accuracy while maintaining data efficiency, such models can be embedded within a $\Delta$-machine learning ($\Delta$-ML)~\cite{Ramakrishnan2015} or multi-fidelity \cite{vinod2025predicting} framework, where a more flexible model is trained on the residuals to higher-level methods.

In this work, we demonstrate the broad applicability of a simple product ansatz model, inspired by Hammett’s original formulation and named the Hammett-Inspired Product (HIP). 
We begin with a brief overview of the HIP algorithm, followed by a detailed description of the datasets used. 
We then present and analyze results across multiple chemical domains, comparing HIP to both in-house ML models and literature benchmarks, and evaluate its performance as a baseline for $\Delta$-ML applications.

\begin{figure*}
    \centering
    \includegraphics[scale=.95]{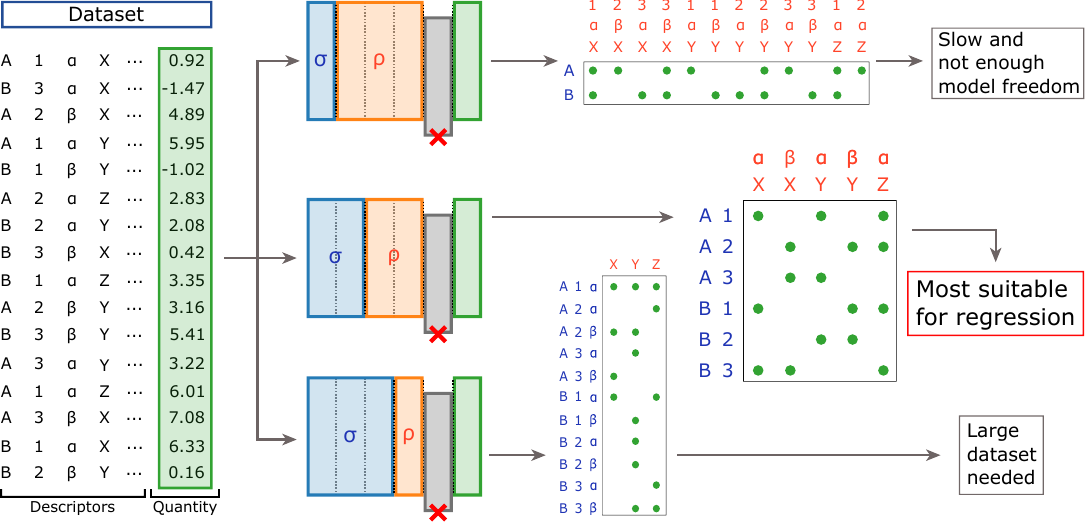}
    \caption{Flowchart of the Hammett approach to an arbitrary dataset. 
    The ellipses refer to descriptors that do not change in the data or are considered irrelevant for the property of interest, and are ignored for this reason. The relevant descriptors for each data point are grouped into two categories so that each point is uniquely described by the combination of two labels. Each set of labels is then paired with a set of $\rho$ or $\sigma$. The resulting matrix-like representation of the dataset is used for the Hammett regression. To get the model with the lowest average prediction errors out of the Hammett regression one has to identify for $\rho$ and $\sigma$ two effective dimensions that are ideally independent (orthogonal) and approximately balanced in the dataset to be used. Examples include solute ($\sigma$) vs.~solvent ($\rho$), or  adsorbate ($\sigma$) vs.~surface ($\rho$).}
    \label{fig:flowchart}
\end{figure*}

\section{Methods and Computational Details}
\subsection{Brief Review of HIP}
To facilitate the discussion, we briefly review the Hammett's equation and summarize how the HIP model emerges. A more comprehensive explanation of the Hammett ansatz for one given property and compound class is also available in our previous work \cite{rakotonirina2024combining}. Further details are also given in the Supplementary Materials (SM). 

The Hammett approach can be viewed as a projection of an observable onto a two-dimensional parameter space, i.e.~the desired property of any target compound is given by the product of two scalars.
For Hammett's original application, the reactions' and substituents' space have been projected to one dimension each.
This generates a Hammett matrix of reaction constants, with a row for each substituent and a column for each reaction.
An indicator of the suitability of a Hammett model to the two-scalar partition is the presence of a linear correlation between the properties measured across different environments or reactions, that is, between any two columns of the Hammett matrix.
The model therefore assumes chemical transferability, namely that trends among functional groups are conserved across different environments.
However, the basis of the Hammett equation is general enough to also be applied to many problems other than reaction constants for a specific chemical subset.
In previous less general and more preliminary work, we described the set of simultaneous regressions making up the model\cite{Bragato2020, rakotonirina2024combining}, and show how to arrive at the product ansatz for metallorganic catalysts\cite{rakotonirina2024combining}.

If each point in the dataset is uniquely identified using two independent variables $\sigma$ and $\rho$ with respective indices, $i$ and $j$, then a Hammett matrix can be built and the target property $P_{ij}$ can be evaluated by a scalar product of $\rho_i$ and $\sigma_j$:

\begin{equation}
    P_{ij} \approx \sigma_i \rho_j 
    \label{eq: Hammett}
\end{equation}
$\sigma$ and $\rho$ are inherently independent and determine the property that is being studied. They must be selected for each specific problem.
An important consequence of using only scalars as parameters is the lack of need to produce an ad-hoc representation for the compounds since any two-label coordinate system is enough to perform the necessary regressions, i.e.~as long as each datapoint can be uniquely identified with two such labels.
Moreover, the 2D projection can effectively remove the need of devising suitable representations tailored different properties: it does not need geometrical or ensemble information to work and all the parameters are obtained via linear regression.

The HIP model accounts for the sets $\{ \rho_j \}$ and $\{\sigma_i \}$ in the following way.
It first evaluates the correlation coefficient $\frac{P_j}{P_{j'}}$, via a Theil-Sen regressor\cite{Theil1950}, between any two columns of the Hammett matrix. 
This constitutes an averaged ratio of the $\rho$ of the two columns given the properties of all compounds recorded across the two.
This translates to $ P_j = \frac{\rho_j}{\rho_{j'}} P_{j'}$ or
\begin{equation}
    \frac{P_j}{P_{j'}} \rho_{j'} - \rho_{j} = 0
    \label{eq:correlation}
\end{equation}
By writing eq.~\ref{eq:correlation} for all pairs of columns, we obtain the overdetermined linear system:
\begin{equation}
    \mathbb{M} \bm{\rho} = \bm{0}
    \label{eq:matr4rho}
\end{equation}
where $\mathbb{M}$ is a matrix containing all the correlation coefficients.
Knowing $\{ \rho_j \}$, each $\sigma_i$ can be obtained by inverting eq.~\ref{eq: Hammett}: $\sigma_i =  P_{ij} / \rho_j$ and averaging over the different columns. 
HIP includes an additional fine-tuning (or balancing) of the values of $\{ \rho_i \}$ via Theil-Sen regression or least-squares following
\begin{equation}
    \rho_j ={\underset {\rho_j}{\mbox{arg min}}}\sum _{i=1}^{m}({\rho_j}\,.\,{\sigma_i}-P_{ij})^{2}
 \label{eq: newrho}
\end{equation}
For numerical reasons, it might be necessary to initially fix one arbitrary reaction constant to 1 in order to avoid trivial solutions when solving eq.~\ref{eq:matr4rho}.

An important difference between the original Hammett model and HIP regards the existence of a reference.
The Hammett equation calibrates the value differences of all reactions relative to a fixed reference reaction.
By contrast, this arbitrary dependence is removed within our HIP model.
While we recognize that Hammett's choice was justified by the model's goal of numerically evaluating the changes a substituent brings, we note that the need for a reference constricts the model's general applicability and, most importantly, introduces a reference bias. 

\subsection{Parameter Selection and Partitioning in HIP}
The Hammett equation is not tied to a specific property, making it broadly applicable to systems that can be described using two effective descriptors. When more than two are present, related features can often be grouped to define the $\rho$ and $\sigma$ axes.
For example, in our previous work on metallorganic complexes, the effects of the ligands were combined into a general parameter $\sigma$~\cite{rakotonirina2024combining}. The most effective partitioning yields a well-populated, balanced descriptor matrix, minimizing extrapolation and reducing prediction error.

To estimate relative $\rho$ values for any two columns of the Hammett matrix, they must share at least two $\sigma$ instances, as implied by Eq.~\ref{eq:correlation}. 
Any strong deviation from the ideal distribution depicted in Figure~\ref{fig:flowchart} will limit the model's freedom and increase the data needed for a quality fit, due to the large number of single linear fits required.
For instance, representing nucleophile–leaving group combinations~\cite{Bragato2020} as a [4$\times$1572] matrix increases average error 4.3-fold compared to a more balanced [12$\times$524] format. This effect is especially relevant in low-data regimes.

HIP also allows irrelevant or redundant descriptors to be omitted without affecting generality (e.g., constant molecular scaffolds across entries, shown as grey columns in Figure~\ref{fig:flowchart}).

If some $\sigma$ values are missing due to incomplete coverage, we apply a simple extrapolation procedure based on known values, using a one-hot encoding of the varying fragment (e.g., substituent, ligand, leaving group). This allows HIP to make reasonable predictions beyond the observed combinatorial space and provides a foundation for future refinements.

\subsection{$\Delta$-ML}
HIP can be combined with more flexible ML models using the $\Delta$-ML approach~\cite{Ramakrishnan2015}. In this framework, the HIP prediction serves as a baseline, and a secondary model is trained to learn the residuals—the difference between the true property and the HIP estimate. This residual surface is typically smoother and less complex, making it easier to learn. In this work, we use Kernel Ridge Regression (KRR) to model the residuals and evaluate the benefit of HIP as a baseline in $\Delta$-ML applications.

\subsection{Datasets}
To gauge the generalizability of Hammett's Ansatz, we have parameterized and applied it to the following data-sets.
\subsubsection{Solvation Free Energies}
We utilized two publicly available datasets, MNSol and Solv@TUM, for experimental free energies of solvation \cite{MNSol,Thompson2004,SolvaTUM}. Due to the challenges associated with accurately modeling hydrogen bonding in water, we excluded it as a solvent from our analysis. 
While this is indeed a significant restriction, we recognize that the modeling of water is itself a complex issue, that has a long history of several methods and ad-hoc solutions\cite{Ouyang2015,Nezbeda2016,Guillot2002,Wallqvist1999}.
After preprocessing, the Solv@TUM dataset comprised 4184 data points from 197 solutes and 70 solvents. We further refined this by excluding solvents with fewer than 30 examples and solutes with fewer than 6 examples to ensure adequate coverage of the combinatorial space (see Figure~\ref{fig:flowchart}). The MNSol subset consisted of 447 data points from 85 solutes and six selected solvents (benzene, carbon tetrachloride, chloroform, cyclohexane, hexadecane, and octanol).

\subsubsection{Formation Energies of Elpasolites}
The elpasolite dataset, generated by Faber et al. \cite{Faber2016}, contains approximately 10'000 quaternary crystal structures made up of main-group elements, computed using density functional theory (DFT) with projector-augmented wave pseudopotentials and the functional of Perdew, Burke, and Ernzerhof, aka PBE \cite{Hohenberg1964,Sham1966,PBE,Blochl1994}. 
The DFT data was used to train ML models of formation energies which were then applied to the remaining possible $\sim$2 million elpasolite candidates. 
In this work, we use the combined dataset to fit the HIP parameters and train the $\Delta$-ML model.
This was done to obtain a fuller Hammett matrix for optimal regression while harnessing the computational efficiency of ML models.

\subsubsection{Adsorption Energies of Small Carbon Species}
Adsorption energy data was obtained from the computational dataset generated by Li et al. \cite{Li2018}, as utilized in the study by Li Xinyu et al. \cite{Li2019}, which includes 272 data points for 68 adsorbates on 4 metal facets.

\subsubsection{HOMO-LUMO gaps of Tungsten Complexes}
For Tungsten-based complexes, we used the dataset reported by Chang et al. \cite{Chang2019}, which provides HOMO-LUMO gaps calculated using DFT at the B3LYP36/Def2SVP level of theory \cite{Beck1993,SVP,Weigend2006}. This dataset consists of 151 complexes of the form $\ce{W(\bond{#}CArR)L_4X}$, where $\ce{Ar}$ is an aryl group, and $\ce{R}$, $\ce{L}$ and $\ce{X}$ are ligands. These 151 entries were generated by systematically combining 5, 6, and 5 different $\ce{R}$, $\ce{L}$, and $\ce{X}$ types, respectively.

\subsubsection{Activation Energies of S$_{\text{N}}$2 Reactions}
To further illustrate the versatility of HIP, we also reproduce our previous successful applications of the model in the context of S$_{\text{N}}$2 \cite{Bragato2020} and C-C cross-coupling (next subsection) reactions \cite{rakotonirina2024combining}.
For the S$_{\text{N}}$2 results, we employed a subset of the QMRxn20 computational dataset \cite{QMRxN} for MP2/6-311G(d) level activation energies of 2433 S$_{\text{N}}$2 reactions involving an ethane backbone (R1R2C1$\ce{\bond{-}}$C2H(R3)(R4)$\ce{\bond{-}}$X). The substituents R1 to R4 included H, NO$_2$, CN, CH$_3$, or NH$_2$, while X, the leaving group or nucleophile, could be H, F, Cl, or Br. A total of 524 substituent combinations were evaluated across the 12 nucleophile-leaving group pairs.

\subsubsection{Binding Energies of Catalysts and Substrates in the Suzuki-Miyaura Reaction}
Lastly, we show the results of the HIP model on the catalyst-substrate binding energies in the oxidative addition of the Suzuki-Miyaura C-C cross-coupling reaction as reported in Ref \cite{rakotonirina2024combining}.
The dataset contained ~25k binding energies for $L_{i}-M_m-L_{j}$ metallorganic catalysts made from combining 6 different metals (Ni,
Pd, Pt, Cu, Ag, and Au) with 91 different ligands (phosphines, N-heterocyclic carbenes, pyridines, and other common ligands).
7054 catalysts were obtained from the publicly available dataset published in Ref. \cite{Meyer2018}.
They were optimized using the AiiDA automated platform \cite{aiida} at the B3LYP-D3/3-21G \cite{B3LYP2010, B3LYP2011} level of theory for the Ni, Pd, Cu, and Ag complexes, and B3LYP-D3/def2-SVP \cite{SVP} for the Pt and Au complexes in Gaussian09 \cite{g09}.
Energies were calculated at the level of B3LYP-D3 / def2-TZVP \cite{SVP}.
The remaining 18062 energies were predicted in our prior work \cite{rakotonirina2024combining} using a KRR model.

\section{Results and Discussion}

\begin{figure*}
	    \centering
    \includegraphics[width=1\textwidth]{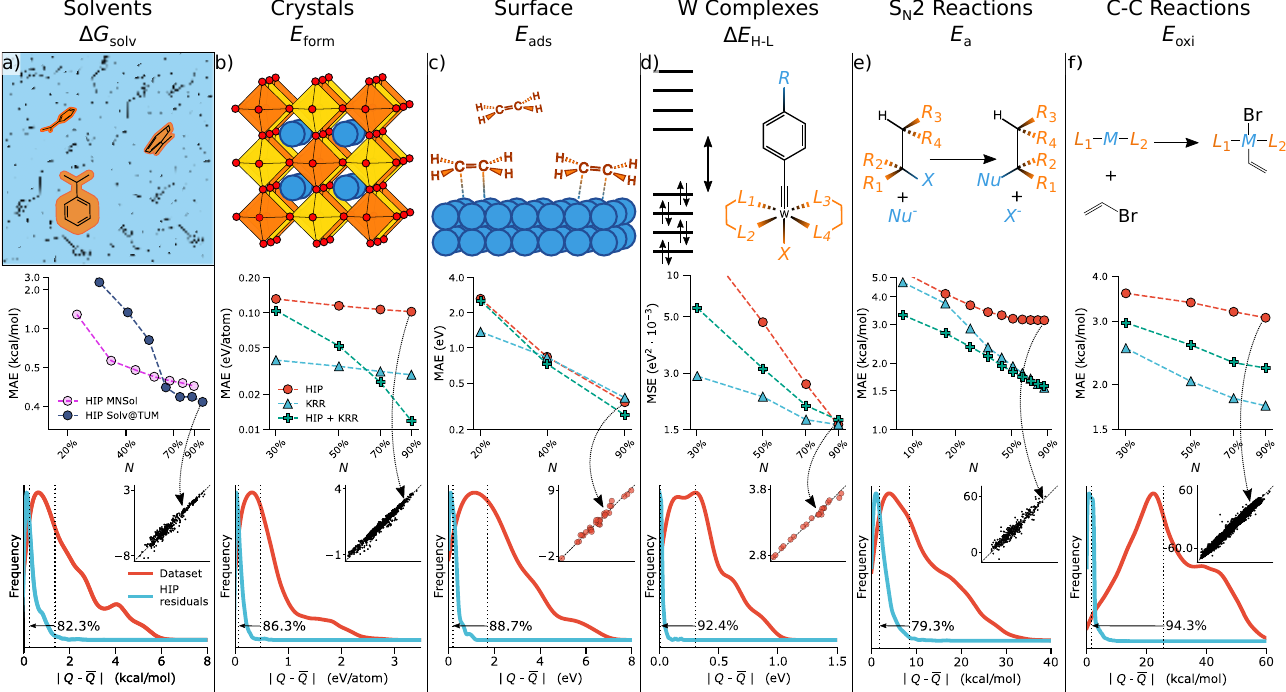}
    \caption{Applications of our Hammett-inspired model to different systems. Top row: schematics of the systems and properties being studied in this work. From left to right: a) Solvation free energies (both experimental~\cite{SolvaTUM} and computational\cite{MNSol}), b) Formation energies of elpasolites\cite{Faber2016}, c) Adsorption energies of small carbon species\cite{Li2019}, d) HOMO-LUMO gaps of metallorganic complexes\cite{Chang2019}, e) Activation Energies for S$_\mathrm{N}$2 reactions\cite{QMRxN}. Middle row: learning curves for our HIP model, a KRR model and a combination of the two (i.e.~$\Delta$-ML), where the KRR model learns the residuals of the HIP regression. Bottom row: the distribution of the absolute deviation from the mean of the dataset and of the Hammett-Inspired Product (HIP) residuals. The dotted vertical lines show the positions of the medians, with the percentage reduction noted on the side. Insets: Scatter plots truth vs.~HIP for the maximum training set sizes considered.}
    \label{fig:complete-pic}
\end{figure*}

In this section, we demonstrate HIP’s performance across the diverse chemical applications outlined earlier. The first row in Figure~\ref{fig:complete-pic} provides an overview of the systems studied, illustrating HIP’s broad applicability.

The second row of Figure~\ref{fig:complete-pic} presents learning curves that assess the model's predictive performance, using the same error metric as in each dataset’s original study to enable direct comparison.
Each point corresponds to the average error from 15- to 25-fold cross-validation runs, plotted against the training set size on a logarithmic scale.
To address statistical variability, a representative example including standard deviation bars is provided for the adsorption system in the SM; the observed trends align with expectations and do not alter the conclusions.
When possible, we included comparisons with a KRR model and a $\Delta$-ML approach using HIP as a baseline. 
All models were trained on the same one-hot-like representations to control for differences in input complexity.


Categorical one-hot encodings were used for the S$_\text{N}$2 reactions, HOMO–LUMO gap prediction, cross-coupling reactions, and catalyst surface in the adsorption problem. Extended-connectivity fingerprints (ECFP)~\cite{ECFP} were used for solvation systems and small carbon adsorbates. For elpasolites, the representation consisted of the group and period indices of each atomic site\cite{Faber2016}.

The bottom row of Figure~\ref{fig:complete-pic} shows how HIP captures linear trends. We compare the distribution of the original target values, centered around their mean, to the distribution of the residuals obtained
after applying the model. The reduction in absolute deviation from the mean quantifies the linear variance explained by HIP, which exceeds 79\% for all systems studied.

Where applicable, we compare HIP to results from alternative models published in the literature.

\subsection{Solvation Free Energies}
Solvated systems, shown in Figure \ref{fig:complete-pic}a, naturally lend themselves to a Hammett-style decomposition, where the solute and solvent define the two axes. We leverage this structure to apply HIP to the prediction of experimental solvation free energies of various solutes in non-aqueous solvents. 
Solvation free energy prediction is a challenging task, often requiring complex representations to capture ensemble behavior~\cite{Hutchinson2019,Riniker2017,Lim2019,Lim2021}. HIP circumvents this step by operating without handcrafted or learned molecular descriptors.

Direct comparison with prior work is limited by differences in dataset composition and inclusion of water, which we exclude, as previously discussed. Therefore, we restrict comparisons to studies using similarly sized, non-aqueous training sets. Table~\ref{table:Solv-compa} summarizes these results.

Despite its simplicity, HIP performs comparably to several non-linear models while using far fewer parameters and significantly less computational effort. For example, Ref.~\cite{Riniker2017} achieves similar errors with subsets of MNSol using LASSO~\cite{LASSO} with 44 descriptors per solvation system, some requiring molecular dynamics simulations. Ref.\cite{Hutchinson2019} uses a one-hot encoding with XGBoost~\cite{xgboost} on MNSol data (octanol and hexadecane), yielding good performance at low-data regimes due to the model complexity. Ref.\cite{Lim2019} reaches similar accuracy with word embeddings and deep neural networks, while Ref.~\cite{Lim2021} applies a 128-dimensional representation on a larger Solv@TUM subset, again with neural networks.

\begin{table}[]
\caption{Prediction root mean square errors in kcal/mol for solvation free energies from literature and HIP. Results correspond to various models based on various data selections (training set size indicated in brackets) extracted from MNSol~\cite{MNSol} and Solv@TUM~\cite{SolvaTUM}.}
\label{table:Solv-compa}
\begin{tabular}{@{}lc}
\toprule
MNSol &\\ \hline
Riniker\cite{Riniker2017}            (83 - 642)     & 0.5 - 1.2        \\
Hutchinson et al.\cite{Hutchinson2019}  (198 - 247)  & 1.19 - 1.65         \\
Delfos\cite{Lim2019}                    (2495)  & 0.57                  \\ 
HIP                            (447)   & 0.72                    \\ \hline \hline
Solv@TUM & \\ \hline
MLSolvA\cite{Lim2021}               (6239)         & 0.42                 \\ 
HIP                            (3765)  & 0.55                     \\ \bottomrule 
\label{table:Solv-compare}
\end{tabular}
\end{table}


\subsection{Formation Energies of Elpasolites}

Unlike solvated systems, 3D crystalline materials lack an obvious two-factor decomposition. For elpasolites (ABC$_2$D$_6$), we propose partitioning the unit cell by isolating an element, e.g., A, as one factor and grouping the remaining sites B, C, and D as the second. This choice is motivated by the large dataset size.

Figure~\ref{fig:complete-pic}b presents HIP’s performance on elpasolite formation energies. As expected for a simple model, HIP underperforms compared to the KRR model from Ref.~\cite{Faber2016}, which achieved 0.1 eV/atom errors at 10k training points. The relatively flat learning curve of HIP can be attributed to the fact that 99.5\% of the HIP training data originated from machine learning based estimates.

Nonetheless, HIP captures meaningful trends: the $\rho$ values learned for each element correlate well with the atomic contributions identified by the atomic representation in Ref.~\cite{Faber2016}. 
A direct comparison of the two approaches is provided in the figure in the SM.
This finding suggests that HIP, despite its simplicity, reflects underlying physical effects influencing formation energy. 
This likely contributes to the remarkably steep learning we observe when using HIP within a $\Delta$-ML framework, where the residuals exhibit clear patterns.

\subsection{Adsorption Energies of Small Carbon Species}

Adsorption naturally lends itself to a two-parameter decomposition: one for the surface and one for the adsorbate. Prior work has established scaling relationships in adsorption processes~\cite{Abild-Pedersen2007,Fernandez2006,Calle2012}, supporting the suitability of a Hammett-style model in this context. Given the central role of adsorption energies in heterogeneous catalysis—especially in constructing volcano plots for catalyst screening~\cite{Gerischer1958,Parsons1958,Norskov2004,Bligaard2004,Norskov2005,Meyer2018}—we applied HIP to the prediction of adsorption energies for small carbon-containing species on four metal surfaces: Cu(111), Pt(111), Pd(111), and Ru(0001). 
Results are shown in Figure~\ref{fig:complete-pic}c. In particular, all models achieve similar overall performance.

The rightmost panel of Figure~\ref{fig:compare-lit} compares our results to those from Ref.\cite{Li2019}, which evaluated multiple representations. 
SOAP was used as a joint representation for both the surface and the adsorbate. 
Alternatively, surface-only representations—Coulomb matrix (CM)\cite{rupp2012coulomb} and elemental properties (EP)\cite{EP}—were combined with one of three adsorbate-specific representations: SLATM\cite{SLATM}, ECFP~\cite{ECFP}, or bag of bonds (BOB)~\cite{BoB}.
HIP rapidly narrows the performance gap with non-linear models as the training set grows. 
This behavior likely reflects improved coverage of the combinatorial space of adsorbate–surface pairs, rather than a simple dependence on the total number of data points.

\subsection{HOMO–LUMO Gaps of Tungsten Complexes}
HIP can be applied at the single-molecule level when distinct molecular substructures, such as ligands, are present. 
We partitioned the ligands into two groups: the $L$ and $X$ ligands were treated as the $\sigma$ parameter, while $R$ was treated as $\rho$. This grouping was found to be the most efficient for capturing systematic trends.

We applied HIP to predict HOMO–LUMO gaps, a key molecular property that governs reactivity, excitation energies, and polarizability. These gaps are especially relevant in the design of catalysts, photovoltaics, and light-emitting diodes~\cite{Kubatkin2003,Roncali2007,Jurow2010,Tao2017,Stoliaroff2020}. Despite their importance, predicting orbital energies remains challenging due to the highly non-monotonic and non-smooth dependence on molecular structure~\cite{mazouin2022selected}.

\begin{figure}[h!]
	    \centering
    \includegraphics[scale=0.9]{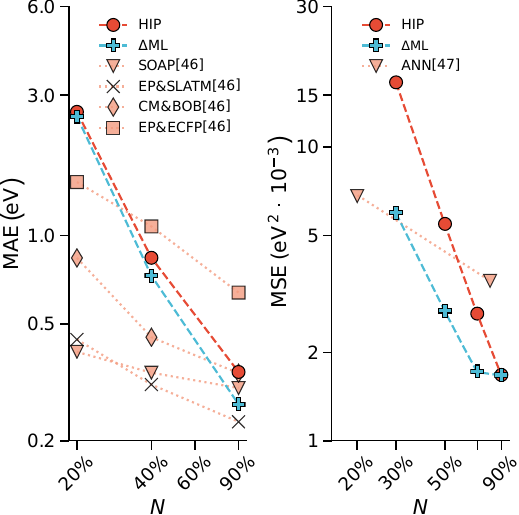}
    \caption{Comparison of HIP and HIP + KRR ($\Delta$-ML) to published ML predictions of the adsorption energies of small carbon species (left), and HOMO-LUMO gaps of metallorganic complexes (right). We re-plot the learning curves from Figure~\ref{fig:complete-pic} for comparison. The learning curve for the HOMO-LUMO gaps is obtained with an artificial neural network for the HOMO-LUMO gaps and the ones for adsorption energies with a KRR model with different representations.}
    \label{fig:compare-lit}
\end{figure}

Nonetheless, Figure~\ref{fig:complete-pic}c shows that much of the variation in this dataset is linear.
HIP starts with a higher error than the non‑linear models, but the gap narrows quickly as more data are supplied. When 90\% of the dataset is used, all three learning curves meet, indicating that the shared representation—not algorithmic complexity—now limits accuracy. HIP, therefore, offers a low‑overhead alternative once an adequate coverage of the combinatorial space is available.
As shown in Figure~\ref{fig:compare-lit} (left), HIP even surpasses the neural network from Ref.~\cite{Chang2019}.
HIP's performance appears to depend more on the completeness of the Hammett matrix than on the total number of data points.
Notably, this is the smallest dataset among those studied, and while small datasets typically yield higher errors, HIP can outperform more complex models once there are enough samples to estimate $\rho$ and $\sigma$. Beyond that point, additional data primarily reinforce the same parameter values, due to their transferability. 

\subsection{Activation Energies of S$_{\text{N}}$2 Reactions}
Activation energies are the original domain of the Hammett equation, whose formulation explicitly assigns two parameters: the reaction type—defined by the nucleophile–leaving group pair—as $\rho$, and the substituents on the molecule as $\sigma$\cite{Hammett1937}. While Hammett’s original study focused on benzene derivatives, we previously demonstrated that HIP can be extended to a broader class of reactions\cite{Bragato2020}.
In column~$e$ of Figure~\ref{fig:complete-pic}, we present results for the prototypical S$_\text{N}$2 reaction, illustrating HIP’s applicability to predicting activation energies in this classical system.

\subsection{Binding Energies of Catalysts and Substrates in the Suzuki-Miyaura Reaction}
An intuitive partitioning into $\rho$ and $\sigma$—assigned to the metal center and ligands, respectively—was straightforward in this case. 
The application HIP is particularly valuable for catalyst discovery, where each experimental or high-level computational evaluation is costly and time-consuming. By enabling rapid estimation of all combinations of metal and ligand parameters, HIP offers an efficient screening tool—particularly valuable in low-data regimes.

As shown in Figure~\ref{fig:complete-pic}, HIP exhibits the highest mean absolute error (MAE), but lags behind kernel ridge regression by only approximately 2 kcal/mol, despite requiring far less training time and memory. Because all models were restricted to a simple one-hot descriptor, KRR is better able to capture residual non-linearities. In contrast, our previous work~\cite{rakotonirina2024combining} used a richer representation, where the $\Delta$-ML model outperformed direct KRR—indicating that the limiting factor is no longer the linear baseline provided by HIP, but the expressiveness of the representation.


\section{Conclusion} 
The product ansatz has a long history in quantum mechanics and physical organic chemistry, from the Hartree product in early quantum chemistry to the linear free energy relationships (LFERs) introduced by Hammett in the late 1930s. Here, we have generalized Hammett’s original product ansatz into what we refer to as the Hammett-Inspired Product (HIP) model, extending it to describe a broad range of chemical properties across diverse regions of chemical compound space.

Our numerical results—spanning solvation energies, crystal formation energies, adsorption energies, HOMO–LUMO gaps, activation energies, and catalyst binding energies—indicate the general applicability of HIP across chemically diverse systems. While results for the latter two were revisited from our earlier studies~\cite{Bragato2020, rakotonirina2024combining}, they have been included here to strengthen the assessment of the model’s transferability. In some cases, HIP performs on par with—or even outperforms—more sophisticated ML models. Together, these examples illustrate how the largely linear structure–property relationships present for many properties in select sub-spaces of chemical space allow HIP to account for over 79\% of the variance across all systems---using minimal input and computational resources.

A major contributor to HIP’s efficiency is its coarse-grained nature and independence from atomistic geometries or ensemble-level information, eliminating the need for complex feature engineering.
When used as a baseline in $\Delta$-ML, the ML model can focus on learning only the residual, higher-order, and non-linear components of the property surface—resulting in improved accuracy with minimal training data.  
Its lack of geometric input and simple linear form enable predictions for tens of thousands of compounds within seconds on a standard CPU. 
Hence, while not a replacement for high-accuracy methods, HIP offers a fast and interpretable baseline for exploring chemical compound space.

Future work may involve extending HIP by incorporating alternative linear free energy parameters, such as Yukawa–Tsuno~\cite{yukawa1959} or Taft descriptors~\cite{Taft1952, Taft1952-2}, to assess their value as enhanced baselines for $\Delta$-ML.

\section*{Acknowledgments}
We acknowledge the support of the Natural Sciences and Engineering Research Council of Canada (NSERC), [funding reference number RGPIN-2023-04853]. Cette recherche a été financée par le Conseil de recherches en sciences naturelles et en génie du Canada (CRSNG), [numéro de référence RGPIN-2023-04853].
This research was undertaken thanks in part to funding provided to the University of Toronto's Acceleration Consortium from the Canada First Research Excellence Fund,
grant number: CFREF-2022-00042.
O.A.v.L. has received support as the Ed Clark Chair of Advanced Materials and as a Canada CIFAR AI Chair.
O.A.v.L. has received funding from the European Research Council (ERC) under the European Union’s Horizon 2020 research and innovation programme (grant agreement No. 772834).
V.D.R. acknowledges support by the Vector Institute's Research Grant.

\bibliography{main.bib}

\begin{thebibliography}{77}%
\makeatletter
\providecommand \@ifxundefined [1]{%
 \@ifx{#1\undefined}
}%
\providecommand \@ifnum [1]{%
 \ifnum #1\expandafter \@firstoftwo
 \else \expandafter \@secondoftwo
 \fi
}%
\providecommand \@ifx [1]{%
 \ifx #1\expandafter \@firstoftwo
 \else \expandafter \@secondoftwo
 \fi
}%
\providecommand \natexlab [1]{#1}%
\providecommand \enquote  [1]{``#1''}%
\providecommand \bibnamefont  [1]{#1}%
\providecommand \bibfnamefont [1]{#1}%
\providecommand \citenamefont [1]{#1}%
\providecommand \href@noop [0]{\@secondoftwo}%
\providecommand \href [0]{\begingroup \@sanitize@url \@href}%
\providecommand \@href[1]{\@@startlink{#1}\@@href}%
\providecommand \@@href[1]{\endgroup#1\@@endlink}%
\providecommand \@sanitize@url [0]{\catcode `\\12\catcode `\$12\catcode
  `\&12\catcode `\#12\catcode `\^12\catcode `\_12\catcode `\%12\relax}%
\providecommand \@@startlink[1]{}%
\providecommand \@@endlink[0]{}%
\providecommand \url  [0]{\begingroup\@sanitize@url \@url }%
\providecommand \@url [1]{\endgroup\@href {#1}{\urlprefix }}%
\providecommand \urlprefix  [0]{URL }%
\providecommand \Eprint [0]{\href }%
\providecommand \doibase [0]{http://dx.doi.org/}%
\providecommand \selectlanguage [0]{\@gobble}%
\providecommand \bibinfo  [0]{\@secondoftwo}%
\providecommand \bibfield  [0]{\@secondoftwo}%
\providecommand \translation [1]{[#1]}%
\providecommand \BibitemOpen [0]{}%
\providecommand \bibitemStop [0]{}%
\providecommand \bibitemNoStop [0]{.\EOS\space}%
\providecommand \EOS [0]{\spacefactor3000\relax}%
\providecommand \BibitemShut  [1]{\csname bibitem#1\endcsname}%
\let\auto@bib@innerbib\@empty
\bibitem [{\citenamefont {Hammett}(1935)}]{Hammett1935}%
  \BibitemOpen
  \bibfield  {author} {\bibinfo {author} {\bibfnamefont {L.~P.}\ \bibnamefont
  {Hammett}},\ }\href {\doibase 10.1021/cr60056a010} {\bibfield  {journal}
  {\bibinfo  {journal} {Chemical Reviews}\ }\textbf {\bibinfo {volume} {17}},\
  \bibinfo {pages} {125} (\bibinfo {year} {1935})}\BibitemShut {NoStop}%
\bibitem [{\citenamefont {Hammett}(1937)}]{Hammett1937}%
  \BibitemOpen
  \bibfield  {author} {\bibinfo {author} {\bibfnamefont {L.~P.}\ \bibnamefont
  {Hammett}},\ }\href {\doibase 10.1021/ja01280a022} {\bibfield  {journal}
  {\bibinfo  {journal} {Journal of the American Chemical Society}\ }\textbf
  {\bibinfo {volume} {59}},\ \bibinfo {pages} {96} (\bibinfo {year}
  {1937})}\BibitemShut {NoStop}%
\bibitem [{\citenamefont {Gerischer}(1958)}]{Gerischer1958}%
  \BibitemOpen
  \bibfield  {author} {\bibinfo {author} {\bibfnamefont {H.}~\bibnamefont
  {Gerischer}},\ }\href@noop {} {\bibfield  {journal} {\bibinfo  {journal}
  {Bulletin des Soci{\'e}t{\'e}s Chimiques Belges}\ }\textbf {\bibinfo {volume}
  {67}},\ \bibinfo {pages} {506} (\bibinfo {year} {1958})}\BibitemShut
  {NoStop}%
\bibitem [{\citenamefont {Parsons}(1958)}]{Parsons1958}%
  \BibitemOpen
  \bibfield  {author} {\bibinfo {author} {\bibfnamefont {R.}~\bibnamefont
  {Parsons}},\ }\href@noop {} {\bibfield  {journal} {\bibinfo  {journal}
  {Transactions of the Faraday Society}\ }\textbf {\bibinfo {volume} {54}},\
  \bibinfo {pages} {1053} (\bibinfo {year} {1958})}\BibitemShut {NoStop}%
\bibitem [{\citenamefont {N{\o}rskov}\ \emph {et~al.}(2004)\citenamefont
  {N{\o}rskov}, \citenamefont {Rossmeisl}, \citenamefont {Logadottir},
  \citenamefont {Lindqvist}, \citenamefont {Kitchin}, \citenamefont
  {Bligaard},\ and\ \citenamefont {Jonsson}}]{Norskov2004}%
  \BibitemOpen
  \bibfield  {author} {\bibinfo {author} {\bibfnamefont {J.~K.}\ \bibnamefont
  {N{\o}rskov}}, \bibinfo {author} {\bibfnamefont {J.}~\bibnamefont
  {Rossmeisl}}, \bibinfo {author} {\bibfnamefont {A.}~\bibnamefont
  {Logadottir}}, \bibinfo {author} {\bibfnamefont {L.}~\bibnamefont
  {Lindqvist}}, \bibinfo {author} {\bibfnamefont {J.~R.}\ \bibnamefont
  {Kitchin}}, \bibinfo {author} {\bibfnamefont {T.}~\bibnamefont {Bligaard}}, \
  and\ \bibinfo {author} {\bibfnamefont {H.}~\bibnamefont {Jonsson}},\
  }\href@noop {} {\bibfield  {journal} {\bibinfo  {journal} {The Journal of
  Physical Chemistry B}\ }\textbf {\bibinfo {volume} {108}},\ \bibinfo {pages}
  {17886} (\bibinfo {year} {2004})}\BibitemShut {NoStop}%
\bibitem [{\citenamefont {Bligaard}\ \emph {et~al.}(2004)\citenamefont
  {Bligaard}, \citenamefont {N{\o}rskov}, \citenamefont {Dahl}, \citenamefont
  {Matthiesen}, \citenamefont {Christensen},\ and\ \citenamefont
  {Sehested}}]{Bligaard2004}%
  \BibitemOpen
  \bibfield  {author} {\bibinfo {author} {\bibfnamefont {T.}~\bibnamefont
  {Bligaard}}, \bibinfo {author} {\bibfnamefont {J.~K.}\ \bibnamefont
  {N{\o}rskov}}, \bibinfo {author} {\bibfnamefont {S.}~\bibnamefont {Dahl}},
  \bibinfo {author} {\bibfnamefont {J.}~\bibnamefont {Matthiesen}}, \bibinfo
  {author} {\bibfnamefont {C.~H.}\ \bibnamefont {Christensen}}, \ and\ \bibinfo
  {author} {\bibfnamefont {J.}~\bibnamefont {Sehested}},\ }\href@noop {}
  {\bibfield  {journal} {\bibinfo  {journal} {Journal of catalysis}\ }\textbf
  {\bibinfo {volume} {224}},\ \bibinfo {pages} {206} (\bibinfo {year}
  {2004})}\BibitemShut {NoStop}%
\bibitem [{\citenamefont {N{\o}rskov}\ \emph {et~al.}(2005)\citenamefont
  {N{\o}rskov}, \citenamefont {Bligaard}, \citenamefont {Logadottir},
  \citenamefont {Kitchin}, \citenamefont {Chen}, \citenamefont {Pandelov},\
  and\ \citenamefont {Stimming}}]{Norskov2005}%
  \BibitemOpen
  \bibfield  {author} {\bibinfo {author} {\bibfnamefont {J.~K.}\ \bibnamefont
  {N{\o}rskov}}, \bibinfo {author} {\bibfnamefont {T.}~\bibnamefont
  {Bligaard}}, \bibinfo {author} {\bibfnamefont {A.}~\bibnamefont
  {Logadottir}}, \bibinfo {author} {\bibfnamefont {J.}~\bibnamefont {Kitchin}},
  \bibinfo {author} {\bibfnamefont {J.~G.}\ \bibnamefont {Chen}}, \bibinfo
  {author} {\bibfnamefont {S.}~\bibnamefont {Pandelov}}, \ and\ \bibinfo
  {author} {\bibfnamefont {U.}~\bibnamefont {Stimming}},\ }\href@noop {}
  {\bibfield  {journal} {\bibinfo  {journal} {Journal of The Electrochemical
  Society}\ }\textbf {\bibinfo {volume} {152}},\ \bibinfo {pages} {J23}
  (\bibinfo {year} {2005})}\BibitemShut {NoStop}%
\bibitem [{\citenamefont {Swain}\ and\ \citenamefont
  {Lupton}(1968)}]{Swain1968}%
  \BibitemOpen
  \bibfield  {author} {\bibinfo {author} {\bibfnamefont {C.~G.}\ \bibnamefont
  {Swain}}\ and\ \bibinfo {author} {\bibfnamefont {E.~C.}\ \bibnamefont
  {Lupton}},\ }\href {\doibase 10.1021/ja01018a024} {\bibfield  {journal}
  {\bibinfo  {journal} {Journal of the American Chemical Society}\ }\textbf
  {\bibinfo {volume} {90}},\ \bibinfo {pages} {4328} (\bibinfo {year}
  {1968})}\BibitemShut {NoStop}%
\bibitem [{\citenamefont {Swain}\ and\ \citenamefont
  {Scott}(1953)}]{Swain1953}%
  \BibitemOpen
  \bibfield  {author} {\bibinfo {author} {\bibfnamefont {C.~G.}\ \bibnamefont
  {Swain}}\ and\ \bibinfo {author} {\bibfnamefont {C.~B.}\ \bibnamefont
  {Scott}},\ }\href@noop {} {\bibfield  {journal} {\bibinfo  {journal} {Journal
  of the American Chemical Society}\ }\textbf {\bibinfo {volume} {75}},\
  \bibinfo {pages} {141} (\bibinfo {year} {1953})}\BibitemShut {NoStop}%
\bibitem [{\citenamefont {Taft~Jr}(1952{\natexlab{a}})}]{Taft1952}%
  \BibitemOpen
  \bibfield  {author} {\bibinfo {author} {\bibfnamefont {R.~W.}\ \bibnamefont
  {Taft~Jr}},\ }\href@noop {} {\bibfield  {journal} {\bibinfo  {journal}
  {Journal of the American Chemical Society}\ }\textbf {\bibinfo {volume}
  {74}},\ \bibinfo {pages} {2729} (\bibinfo {year}
  {1952}{\natexlab{a}})}\BibitemShut {NoStop}%
\bibitem [{\citenamefont {Taft~Jr}(1952{\natexlab{b}})}]{Taft1952-2}%
  \BibitemOpen
  \bibfield  {author} {\bibinfo {author} {\bibfnamefont {R.~W.}\ \bibnamefont
  {Taft~Jr}},\ }\href@noop {} {\bibfield  {journal} {\bibinfo  {journal}
  {Journal of the American Chemical Society}\ }\textbf {\bibinfo {volume}
  {74}},\ \bibinfo {pages} {3120} (\bibinfo {year}
  {1952}{\natexlab{b}})}\BibitemShut {NoStop}%
\bibitem [{\citenamefont {Taft~Jr}(1953)}]{Taft1953}%
  \BibitemOpen
  \bibfield  {author} {\bibinfo {author} {\bibfnamefont {R.~W.}\ \bibnamefont
  {Taft~Jr}},\ }\href@noop {} {\bibfield  {journal} {\bibinfo  {journal}
  {Journal of the American Chemical Society}\ }\textbf {\bibinfo {volume}
  {75}},\ \bibinfo {pages} {4538} (\bibinfo {year} {1953})}\BibitemShut
  {NoStop}%
\bibitem [{\citenamefont {Santiago}\ \emph {et~al.}(2016)\citenamefont
  {Santiago}, \citenamefont {Milo},\ and\ \citenamefont
  {Sigman}}]{Santiago2016}%
  \BibitemOpen
  \bibfield  {author} {\bibinfo {author} {\bibfnamefont {C.~B.}\ \bibnamefont
  {Santiago}}, \bibinfo {author} {\bibfnamefont {A.}~\bibnamefont {Milo}}, \
  and\ \bibinfo {author} {\bibfnamefont {M.~S.}\ \bibnamefont {Sigman}},\
  }\href {\doibase 10.1021/jacs.6b08799} {\bibfield  {journal} {\bibinfo
  {journal} {Journal of the American Chemical Society}\ }\textbf {\bibinfo
  {volume} {138}},\ \bibinfo {pages} {13424} (\bibinfo {year}
  {2016})}\BibitemShut {NoStop}%
\bibitem [{\citenamefont {Jahagirdar}\ \emph {et~al.}(1988)\citenamefont
  {Jahagirdar}, \citenamefont {Arbad},\ and\ \citenamefont
  {Kharwadkar}}]{Jahagirdar1988}%
  \BibitemOpen
  \bibfield  {author} {\bibinfo {author} {\bibfnamefont {D.}~\bibnamefont
  {Jahagirdar}}, \bibinfo {author} {\bibfnamefont {B.}~\bibnamefont {Arbad}}, \
  and\ \bibinfo {author} {\bibfnamefont {R.}~\bibnamefont {Kharwadkar}},\
  }\href@noop {} {\bibfield  {journal} {\bibinfo  {journal} {Indian Journal of
  Chemistry}\ }\textbf {\bibinfo {volume} {27A}},\ \bibinfo {pages} {601}
  (\bibinfo {year} {1988})}\BibitemShut {NoStop}%
\bibitem [{\citenamefont {Kondo}\ \emph {et~al.}(1969)\citenamefont {Kondo},
  \citenamefont {Matsui},\ and\ \citenamefont {Tokura}}]{Kondo1969}%
  \BibitemOpen
  \bibfield  {author} {\bibinfo {author} {\bibfnamefont {Y.}~\bibnamefont
  {Kondo}}, \bibinfo {author} {\bibfnamefont {T.}~\bibnamefont {Matsui}}, \
  and\ \bibinfo {author} {\bibfnamefont {N.}~\bibnamefont {Tokura}},\
  }\href@noop {} {\bibfield  {journal} {\bibinfo  {journal} {Bulletin of the
  Chemical Society of Japan}\ }\textbf {\bibinfo {volume} {42}},\ \bibinfo
  {pages} {1037} (\bibinfo {year} {1969})}\BibitemShut {NoStop}%
\bibitem [{\citenamefont {Grunwald}\ and\ \citenamefont
  {Winstein}(1948)}]{Grunwald1948}%
  \BibitemOpen
  \bibfield  {author} {\bibinfo {author} {\bibfnamefont {E.}~\bibnamefont
  {Grunwald}}\ and\ \bibinfo {author} {\bibfnamefont {S.}~\bibnamefont
  {Winstein}},\ }\href {\doibase 10.1021/ja01182a117} {\bibfield  {journal}
  {\bibinfo  {journal} {Journal of the American Chemical Society}\ }\textbf
  {\bibinfo {volume} {70}},\ \bibinfo {pages} {846} (\bibinfo {year}
  {1948})}\BibitemShut {NoStop}%
\bibitem [{\citenamefont {Winstein}\ \emph {et~al.}(1951)\citenamefont
  {Winstein}, \citenamefont {Grunwald},\ and\ \citenamefont
  {Jones}}]{Winstein1951}%
  \BibitemOpen
  \bibfield  {author} {\bibinfo {author} {\bibfnamefont {S.}~\bibnamefont
  {Winstein}}, \bibinfo {author} {\bibfnamefont {E.}~\bibnamefont {Grunwald}},
  \ and\ \bibinfo {author} {\bibfnamefont {H.~W.}\ \bibnamefont {Jones}},\
  }\href {\doibase 10.1021/ja01150a078} {\bibfield  {journal} {\bibinfo
  {journal} {Journal of the American Chemical Society}\ }\textbf {\bibinfo
  {volume} {73}},\ \bibinfo {pages} {2700} (\bibinfo {year}
  {1951})}\BibitemShut {NoStop}%
\bibitem [{\citenamefont {Dobrowolski}\ \emph {et~al.}(2018)\citenamefont
  {Dobrowolski}, \citenamefont {Lipi{\'n}ski},\ and\ \citenamefont
  {Karpi{\'n}ska}}]{Dobrowolski2018}%
  \BibitemOpen
  \bibfield  {author} {\bibinfo {author} {\bibfnamefont {J.~C.}\ \bibnamefont
  {Dobrowolski}}, \bibinfo {author} {\bibfnamefont {P.~F.~J.}\ \bibnamefont
  {Lipi{\'n}ski}}, \ and\ \bibinfo {author} {\bibfnamefont {G.}~\bibnamefont
  {Karpi{\'n}ska}},\ }\href@noop {} {\bibfield  {journal} {\bibinfo  {journal}
  {The Journal of Physical Chemistry A}\ }\textbf {\bibinfo {volume} {122}},\
  \bibinfo {pages} {4609} (\bibinfo {year} {2018})}\BibitemShut {NoStop}%
\bibitem [{\citenamefont {Song}\ \emph {et~al.}(2006)\citenamefont {Song},
  \citenamefont {Zapata},\ and\ \citenamefont {Eng}}]{Song2006}%
  \BibitemOpen
  \bibfield  {author} {\bibinfo {author} {\bibfnamefont {X.}~\bibnamefont
  {Song}}, \bibinfo {author} {\bibfnamefont {A.}~\bibnamefont {Zapata}}, \ and\
  \bibinfo {author} {\bibfnamefont {G.}~\bibnamefont {Eng}},\ }\href {\doibase
  10.1016/j.jorganchem.2005.12.003} {\bibfield  {journal} {\bibinfo  {journal}
  {Journal of Organometallic Chemistry}\ }\textbf {\bibinfo {volume} {691}},\
  \bibinfo {pages} {1756} (\bibinfo {year} {2006})}\BibitemShut {NoStop}%
\bibitem [{\citenamefont {Masui}\ and\ \citenamefont
  {Lever}(1993)}]{Masui1993}%
  \BibitemOpen
  \bibfield  {author} {\bibinfo {author} {\bibfnamefont {H.}~\bibnamefont
  {Masui}}\ and\ \bibinfo {author} {\bibfnamefont {A.~B.~P.}\ \bibnamefont
  {Lever}},\ }\href {\doibase 10.1021/ic00062a052} {\bibfield  {journal}
  {\bibinfo  {journal} {Inorganic Chemistry}\ }\textbf {\bibinfo {volume}
  {32}},\ \bibinfo {pages} {2199} (\bibinfo {year} {1993})}\BibitemShut
  {NoStop}%
\bibitem [{\citenamefont {Buszta}\ \emph {et~al.}(2019)\citenamefont {Buszta},
  \citenamefont {Depa}, \citenamefont {Bajek},\ and\ \citenamefont
  {Groszek}}]{Buszta2019}%
  \BibitemOpen
  \bibfield  {author} {\bibinfo {author} {\bibfnamefont {N.}~\bibnamefont
  {Buszta}}, \bibinfo {author} {\bibfnamefont {W.~J.}\ \bibnamefont {Depa}},
  \bibinfo {author} {\bibfnamefont {A.}~\bibnamefont {Bajek}}, \ and\ \bibinfo
  {author} {\bibfnamefont {G.}~\bibnamefont {Groszek}},\ }\href@noop {}
  {\bibfield  {journal} {\bibinfo  {journal} {Chemical Papers}\ }\textbf
  {\bibinfo {volume} {73}},\ \bibinfo {pages} {2885} (\bibinfo {year}
  {2019})}\BibitemShut {NoStop}%
\bibitem [{\citenamefont {Kumar}\ \emph {et~al.}(2018)\citenamefont {Kumar},
  \citenamefont {Tibbitts}, \citenamefont {Newell}, \citenamefont {Panthi},
  \citenamefont {Mukhopadhyay}, \citenamefont {Rioux}, \citenamefont {Pursell},
  \citenamefont {Janik},\ and\ \citenamefont {Chandler}}]{Kumar2018}%
  \BibitemOpen
  \bibfield  {author} {\bibinfo {author} {\bibfnamefont {G.}~\bibnamefont
  {Kumar}}, \bibinfo {author} {\bibfnamefont {L.}~\bibnamefont {Tibbitts}},
  \bibinfo {author} {\bibfnamefont {J.}~\bibnamefont {Newell}}, \bibinfo
  {author} {\bibfnamefont {B.}~\bibnamefont {Panthi}}, \bibinfo {author}
  {\bibfnamefont {A.}~\bibnamefont {Mukhopadhyay}}, \bibinfo {author}
  {\bibfnamefont {R.~M.}\ \bibnamefont {Rioux}}, \bibinfo {author}
  {\bibfnamefont {C.~J.}\ \bibnamefont {Pursell}}, \bibinfo {author}
  {\bibfnamefont {M.}~\bibnamefont {Janik}}, \ and\ \bibinfo {author}
  {\bibfnamefont {B.~D.}\ \bibnamefont {Chandler}},\ }\href@noop {} {\bibfield
  {journal} {\bibinfo  {journal} {Nature Chemistry}\ }\textbf {\bibinfo
  {volume} {10}},\ \bibinfo {pages} {268} (\bibinfo {year} {2018})}\BibitemShut
  {NoStop}%
\bibitem [{\citenamefont {Venkataraman}\ \emph {et~al.}(2007)\citenamefont
  {Venkataraman}, \citenamefont {Park}, \citenamefont {Whalley}, \citenamefont
  {Nuckolls}, \citenamefont {Hybertsen},\ and\ \citenamefont
  {Steigerwald}}]{Venkataraman2007}%
  \BibitemOpen
  \bibfield  {author} {\bibinfo {author} {\bibfnamefont {L.}~\bibnamefont
  {Venkataraman}}, \bibinfo {author} {\bibfnamefont {Y.~S.}\ \bibnamefont
  {Park}}, \bibinfo {author} {\bibfnamefont {A.~C.}\ \bibnamefont {Whalley}},
  \bibinfo {author} {\bibfnamefont {C.}~\bibnamefont {Nuckolls}}, \bibinfo
  {author} {\bibfnamefont {M.~S.}\ \bibnamefont {Hybertsen}}, \ and\ \bibinfo
  {author} {\bibfnamefont {M.~L.}\ \bibnamefont {Steigerwald}},\ }\href@noop {}
  {\bibfield  {journal} {\bibinfo  {journal} {Nano Letters}\ }\textbf {\bibinfo
  {volume} {7}},\ \bibinfo {pages} {502} (\bibinfo {year} {2007})}\BibitemShut
  {NoStop}%
\bibitem [{\citenamefont {Bragato}\ \emph {et~al.}(2020)\citenamefont
  {Bragato}, \citenamefont {von Rudorff},\ and\ \citenamefont {von
  Lilienfeld}}]{Bragato2020}%
  \BibitemOpen
  \bibfield  {author} {\bibinfo {author} {\bibfnamefont {M.}~\bibnamefont
  {Bragato}}, \bibinfo {author} {\bibfnamefont {G.~F.}\ \bibnamefont {von
  Rudorff}}, \ and\ \bibinfo {author} {\bibfnamefont {O.~A.}\ \bibnamefont {von
  Lilienfeld}},\ }\href@noop {} {\bibfield  {journal} {\bibinfo  {journal}
  {Chemical Science}\ }\textbf {\bibinfo {volume} {11}},\ \bibinfo {pages}
  {11859} (\bibinfo {year} {2020})}\BibitemShut {NoStop}%
\bibitem [{\citenamefont {von Rudorff}\ and\ \citenamefont {von
  Lilienfeld}(2019)}]{vonRudorff2019}%
  \BibitemOpen
  \bibfield  {author} {\bibinfo {author} {\bibfnamefont {G.~F.}\ \bibnamefont
  {von Rudorff}}\ and\ \bibinfo {author} {\bibfnamefont {O.~A.}\ \bibnamefont
  {von Lilienfeld}},\ }\href@noop {} {\bibfield  {journal} {\bibinfo  {journal}
  {The Journal of Physical Chemistry B}\ }\textbf {\bibinfo {volume} {123}},\
  \bibinfo {pages} {10073} (\bibinfo {year} {2019})}\BibitemShut {NoStop}%
\bibitem [{\citenamefont {von Rudorff}\ and\ \citenamefont {von
  Lilienfeld}(2020)}]{vonRudorff2020}%
  \BibitemOpen
  \bibfield  {author} {\bibinfo {author} {\bibfnamefont {G.~F.}\ \bibnamefont
  {von Rudorff}}\ and\ \bibinfo {author} {\bibfnamefont {O.~A.}\ \bibnamefont
  {von Lilienfeld}},\ }\href@noop {} {\bibfield  {journal} {\bibinfo  {journal}
  {Physical Review Research}\ }\textbf {\bibinfo {volume} {2}},\ \bibinfo
  {pages} {023220} (\bibinfo {year} {2020})}\BibitemShut {NoStop}%
\bibitem [{\citenamefont {von Rudorff}\ and\ \citenamefont {von
  Lilienfeld}(2021)}]{vonRudorff2021}%
  \BibitemOpen
  \bibfield  {author} {\bibinfo {author} {\bibfnamefont {G.~F.}\ \bibnamefont
  {von Rudorff}}\ and\ \bibinfo {author} {\bibfnamefont {O.~A.}\ \bibnamefont
  {von Lilienfeld}},\ }\href@noop {} {\bibfield  {journal} {\bibinfo  {journal}
  {Science Advances}\ }\textbf {\bibinfo {volume} {7}},\ \bibinfo {pages}
  {eabf1173} (\bibinfo {year} {2021})}\BibitemShut {NoStop}%
\bibitem [{\citenamefont {Isayev}\ \emph {et~al.}(2017)\citenamefont {Isayev},
  \citenamefont {Oses}, \citenamefont {Toher}, \citenamefont {Gossett},
  \citenamefont {Curtarolo},\ and\ \citenamefont
  {Tropsha}}]{isayev2017universal}%
  \BibitemOpen
  \bibfield  {author} {\bibinfo {author} {\bibfnamefont {O.}~\bibnamefont
  {Isayev}}, \bibinfo {author} {\bibfnamefont {C.}~\bibnamefont {Oses}},
  \bibinfo {author} {\bibfnamefont {C.}~\bibnamefont {Toher}}, \bibinfo
  {author} {\bibfnamefont {E.}~\bibnamefont {Gossett}}, \bibinfo {author}
  {\bibfnamefont {S.}~\bibnamefont {Curtarolo}}, \ and\ \bibinfo {author}
  {\bibfnamefont {A.}~\bibnamefont {Tropsha}},\ }\href@noop {} {\bibfield
  {journal} {\bibinfo  {journal} {Nature Communications}\ }\textbf {\bibinfo
  {volume} {8}},\ \bibinfo {pages} {15679} (\bibinfo {year}
  {2017})}\BibitemShut {NoStop}%
\bibitem [{\citenamefont {Ramakrishnan}\ \emph {et~al.}(2015)\citenamefont
  {Ramakrishnan}, \citenamefont {Dral}, \citenamefont {Rupp},\ and\
  \citenamefont {von Lilienfeld}}]{Ramakrishnan2015}%
  \BibitemOpen
  \bibfield  {author} {\bibinfo {author} {\bibfnamefont {R.}~\bibnamefont
  {Ramakrishnan}}, \bibinfo {author} {\bibfnamefont {P.~O.}\ \bibnamefont
  {Dral}}, \bibinfo {author} {\bibfnamefont {M.}~\bibnamefont {Rupp}}, \ and\
  \bibinfo {author} {\bibfnamefont {O.~A.}\ \bibnamefont {von Lilienfeld}},\
  }\href {\doibase 10.1021/acs.jctc.5b00099} {\bibfield  {journal} {\bibinfo
  {journal} {Journal of Chemical Theory and Computation}\ }\textbf {\bibinfo
  {volume} {11}},\ \bibinfo {pages} {2087} (\bibinfo {year} {2015})},\ \bibinfo
  {note} {pMID: 26574412}\BibitemShut {NoStop}%
\bibitem [{\citenamefont {Vinod}\ \emph {et~al.}(2025)\citenamefont {Vinod},
  \citenamefont {Lyu}, \citenamefont {Ruth}, \citenamefont {R.~Schreiner},
  \citenamefont {Kleinekath{\"o}fer},\ and\ \citenamefont
  {Zaspel}}]{vinod2025predicting}%
  \BibitemOpen
  \bibfield  {author} {\bibinfo {author} {\bibfnamefont {V.}~\bibnamefont
  {Vinod}}, \bibinfo {author} {\bibfnamefont {D.}~\bibnamefont {Lyu}}, \bibinfo
  {author} {\bibfnamefont {M.}~\bibnamefont {Ruth}}, \bibinfo {author}
  {\bibfnamefont {P.}~\bibnamefont {R.~Schreiner}}, \bibinfo {author}
  {\bibfnamefont {U.}~\bibnamefont {Kleinekath{\"o}fer}}, \ and\ \bibinfo
  {author} {\bibfnamefont {P.}~\bibnamefont {Zaspel}},\ }\href@noop {}
  {\bibfield  {journal} {\bibinfo  {journal} {Journal of Computational
  Chemistry}\ }\textbf {\bibinfo {volume} {46}},\ \bibinfo {pages} {e70056}
  (\bibinfo {year} {2025})}\BibitemShut {NoStop}%
\bibitem [{\citenamefont {Rakotonirina}\ \emph {et~al.}(2024)\citenamefont
  {Rakotonirina}, \citenamefont {Bragato}, \citenamefont {Heinen},\ and\
  \citenamefont {von Lilienfeld}}]{rakotonirina2024combining}%
  \BibitemOpen
  \bibfield  {author} {\bibinfo {author} {\bibfnamefont {V.~D.}\ \bibnamefont
  {Rakotonirina}}, \bibinfo {author} {\bibfnamefont {M.}~\bibnamefont
  {Bragato}}, \bibinfo {author} {\bibfnamefont {S.}~\bibnamefont {Heinen}}, \
  and\ \bibinfo {author} {\bibfnamefont {O.~A.}\ \bibnamefont {von
  Lilienfeld}},\ }\href {\doibase 10.1039/D4DD00228H} {\bibfield  {journal}
  {\bibinfo  {journal} {Digital Discovery}\ }\textbf {\bibinfo {volume} {3}},\
  \bibinfo {pages} {2487} (\bibinfo {year} {2024})}\BibitemShut {NoStop}%
\bibitem [{\citenamefont {Theil}(1950)}]{Theil1950}%
  \BibitemOpen
  \bibfield  {author} {\bibinfo {author} {\bibfnamefont {H.}~\bibnamefont
  {Theil}},\ }\href@noop {} {\bibfield  {journal} {\bibinfo  {journal}
  {Indagationes mathematicae}\ }\textbf {\bibinfo {volume} {12}},\ \bibinfo
  {pages} {173} (\bibinfo {year} {1950})}\BibitemShut {NoStop}%
\bibitem [{\citenamefont {Marenich}\ \emph {et~al.}(2020)\citenamefont
  {Marenich}, \citenamefont {Kelly}, \citenamefont {Thompson}, \citenamefont
  {Hawkins}, \citenamefont {Chambers}, \citenamefont {Giesen}, \citenamefont
  {Winget}, \citenamefont {Cramer},\ and\ \citenamefont {Truhlar}}]{MNSol}%
  \BibitemOpen
  \bibfield  {author} {\bibinfo {author} {\bibfnamefont {A.~V.}\ \bibnamefont
  {Marenich}}, \bibinfo {author} {\bibfnamefont {C.~P.}\ \bibnamefont {Kelly}},
  \bibinfo {author} {\bibfnamefont {J.~D.}\ \bibnamefont {Thompson}}, \bibinfo
  {author} {\bibfnamefont {G.~D.}\ \bibnamefont {Hawkins}}, \bibinfo {author}
  {\bibfnamefont {C.~C.}\ \bibnamefont {Chambers}}, \bibinfo {author}
  {\bibfnamefont {D.~J.}\ \bibnamefont {Giesen}}, \bibinfo {author}
  {\bibfnamefont {P.}~\bibnamefont {Winget}}, \bibinfo {author} {\bibfnamefont
  {C.~J.}\ \bibnamefont {Cramer}}, \ and\ \bibinfo {author} {\bibfnamefont
  {D.~G.}\ \bibnamefont {Truhlar}},\ }\href {\doibase
  https://doi.org/10.13020/3eks-j059} {\enquote {\bibinfo {title} {Minnesota
  {S}olvation {D}atabase ({MNSOL}) version 2012.}}\ } (\bibinfo {year}
  {2020})\BibitemShut {NoStop}%
\bibitem [{\citenamefont {Thompson}\ \emph {et~al.}(2004)\citenamefont
  {Thompson}, \citenamefont {Cramer},\ and\ \citenamefont
  {Truhlar}}]{Thompson2004}%
  \BibitemOpen
  \bibfield  {author} {\bibinfo {author} {\bibfnamefont {J.~D.}\ \bibnamefont
  {Thompson}}, \bibinfo {author} {\bibfnamefont {C.~J.}\ \bibnamefont
  {Cramer}}, \ and\ \bibinfo {author} {\bibfnamefont {D.~G.}\ \bibnamefont
  {Truhlar}},\ }\href@noop {} {\bibfield  {journal} {\bibinfo  {journal} {The
  Journal of Physical Chemistry A}\ }\textbf {\bibinfo {volume} {108}},\
  \bibinfo {pages} {6532} (\bibinfo {year} {2004})}\BibitemShut {NoStop}%
\bibitem [{\citenamefont {Hille}\ \emph {et~al.}(2018)\citenamefont {Hille},
  \citenamefont {Deimel}, \citenamefont {Kunkel}, \citenamefont {Acree},
  \citenamefont {Reuter},\ and\ \citenamefont {Oberhofer}}]{SolvaTUM}%
  \BibitemOpen
  \bibfield  {author} {\bibinfo {author} {\bibfnamefont {S.}~\bibnamefont
  {Hille}, \bibfnamefont {Christophand~Ringe}}, \bibinfo {author}
  {\bibfnamefont {M.}~\bibnamefont {Deimel}}, \bibinfo {author} {\bibfnamefont
  {C.}~\bibnamefont {Kunkel}}, \bibinfo {author} {\bibfnamefont {W.~E.}\
  \bibnamefont {Acree}}, \bibinfo {author} {\bibfnamefont {K.}~\bibnamefont
  {Reuter}}, \ and\ \bibinfo {author} {\bibfnamefont {H.}~\bibnamefont
  {Oberhofer}},\ }\href {\doibase 10.14459/2018mp1452571.001} {\enquote
  {\bibinfo {title} {Solv@tum v 1.0},}\ } (\bibinfo {year} {2018})\BibitemShut
  {NoStop}%
\bibitem [{\citenamefont {Ouyang}\ and\ \citenamefont
  {Bettens}(2015)}]{Ouyang2015}%
  \BibitemOpen
  \bibfield  {author} {\bibinfo {author} {\bibfnamefont {J.~F.}\ \bibnamefont
  {Ouyang}}\ and\ \bibinfo {author} {\bibfnamefont {R.}~\bibnamefont
  {Bettens}},\ }\href@noop {} {\bibfield  {journal} {\bibinfo  {journal}
  {CHIMIA International Journal for Chemistry}\ }\textbf {\bibinfo {volume}
  {69}},\ \bibinfo {pages} {104} (\bibinfo {year} {2015})}\BibitemShut
  {NoStop}%
\bibitem [{\citenamefont {Nezbeda}\ \emph {et~al.}(2016)\citenamefont
  {Nezbeda}, \citenamefont {Mou{\v{c}}ka},\ and\ \citenamefont
  {Smith}}]{Nezbeda2016}%
  \BibitemOpen
  \bibfield  {author} {\bibinfo {author} {\bibfnamefont {I.}~\bibnamefont
  {Nezbeda}}, \bibinfo {author} {\bibfnamefont {F.}~\bibnamefont
  {Mou{\v{c}}ka}}, \ and\ \bibinfo {author} {\bibfnamefont {W.~R.}\
  \bibnamefont {Smith}},\ }\href@noop {} {\bibfield  {journal} {\bibinfo
  {journal} {Molecular Physics}\ }\textbf {\bibinfo {volume} {114}},\ \bibinfo
  {pages} {1665} (\bibinfo {year} {2016})}\BibitemShut {NoStop}%
\bibitem [{\citenamefont {Guillot}(2002)}]{Guillot2002}%
  \BibitemOpen
  \bibfield  {author} {\bibinfo {author} {\bibfnamefont {B.}~\bibnamefont
  {Guillot}},\ }\href@noop {} {\bibfield  {journal} {\bibinfo  {journal}
  {Journal of Molecular Liquids}\ }\textbf {\bibinfo {volume} {101}},\ \bibinfo
  {pages} {219} (\bibinfo {year} {2002})}\BibitemShut {NoStop}%
\bibitem [{\citenamefont {Wallqvist}\ and\ \citenamefont
  {Mountain}(1999)}]{Wallqvist1999}%
  \BibitemOpen
  \bibfield  {author} {\bibinfo {author} {\bibfnamefont {A.}~\bibnamefont
  {Wallqvist}}\ and\ \bibinfo {author} {\bibfnamefont {R.~D.}\ \bibnamefont
  {Mountain}},\ }\href@noop {} {\bibfield  {journal} {\bibinfo  {journal}
  {Reviews in Computational Chemistry}\ ,\ \bibinfo {pages} {183}} (\bibinfo
  {year} {1999})}\BibitemShut {NoStop}%
\bibitem [{\citenamefont {Faber}\ \emph {et~al.}(2016)\citenamefont {Faber},
  \citenamefont {Lindmaa}, \citenamefont {Von~Lilienfeld},\ and\ \citenamefont
  {Armiento}}]{Faber2016}%
  \BibitemOpen
  \bibfield  {author} {\bibinfo {author} {\bibfnamefont {F.~A.}\ \bibnamefont
  {Faber}}, \bibinfo {author} {\bibfnamefont {A.}~\bibnamefont {Lindmaa}},
  \bibinfo {author} {\bibfnamefont {O.~A.}\ \bibnamefont {Von~Lilienfeld}}, \
  and\ \bibinfo {author} {\bibfnamefont {R.}~\bibnamefont {Armiento}},\
  }\href@noop {} {\bibfield  {journal} {\bibinfo  {journal} {Physical Review
  Letters}\ }\textbf {\bibinfo {volume} {117}},\ \bibinfo {pages} {135502}
  (\bibinfo {year} {2016})}\BibitemShut {NoStop}%
\bibitem [{\citenamefont {Hohenberg}\ and\ \citenamefont
  {Kohn}(1964)}]{Hohenberg1964}%
  \BibitemOpen
  \bibfield  {author} {\bibinfo {author} {\bibfnamefont {P.}~\bibnamefont
  {Hohenberg}}\ and\ \bibinfo {author} {\bibfnamefont {W.}~\bibnamefont
  {Kohn}},\ }\href@noop {} {\bibfield  {journal} {\bibinfo  {journal} {Physical
  Review}\ }\textbf {\bibinfo {volume} {136}},\ \bibinfo {pages} {B864}
  (\bibinfo {year} {1964})}\BibitemShut {NoStop}%
\bibitem [{\citenamefont {Sham}\ and\ \citenamefont {Kohn}(1966)}]{Sham1966}%
  \BibitemOpen
  \bibfield  {author} {\bibinfo {author} {\bibfnamefont {L.~J.}\ \bibnamefont
  {Sham}}\ and\ \bibinfo {author} {\bibfnamefont {W.}~\bibnamefont {Kohn}},\
  }\href@noop {} {\bibfield  {journal} {\bibinfo  {journal} {Physical Review}\
  }\textbf {\bibinfo {volume} {145}},\ \bibinfo {pages} {561} (\bibinfo {year}
  {1966})}\BibitemShut {NoStop}%
\bibitem [{\citenamefont {Perdew}\ \emph {et~al.}(1996)\citenamefont {Perdew},
  \citenamefont {Burke},\ and\ \citenamefont {Ernzerhof}}]{PBE}%
  \BibitemOpen
  \bibfield  {author} {\bibinfo {author} {\bibfnamefont {J.~P.}\ \bibnamefont
  {Perdew}}, \bibinfo {author} {\bibfnamefont {K.}~\bibnamefont {Burke}}, \
  and\ \bibinfo {author} {\bibfnamefont {M.}~\bibnamefont {Ernzerhof}},\
  }\href@noop {} {\bibfield  {journal} {\bibinfo  {journal} {Physical Review
  Letters}\ }\textbf {\bibinfo {volume} {77}},\ \bibinfo {pages} {3865}
  (\bibinfo {year} {1996})}\BibitemShut {NoStop}%
\bibitem [{\citenamefont {Bl{\"o}chl}(1994)}]{Blochl1994}%
  \BibitemOpen
  \bibfield  {author} {\bibinfo {author} {\bibfnamefont {P.~E.}\ \bibnamefont
  {Bl{\"o}chl}},\ }\href@noop {} {\bibfield  {journal} {\bibinfo  {journal}
  {Physical Review B}\ }\textbf {\bibinfo {volume} {50}},\ \bibinfo {pages}
  {17953} (\bibinfo {year} {1994})}\BibitemShut {NoStop}%
\bibitem [{\citenamefont {Li}\ \emph {et~al.}(2018)\citenamefont {Li},
  \citenamefont {Garc{\'\i}a-Muelas},\ and\ \citenamefont
  {L{\'o}pez}}]{Li2018}%
  \BibitemOpen
  \bibfield  {author} {\bibinfo {author} {\bibfnamefont {Q.}~\bibnamefont
  {Li}}, \bibinfo {author} {\bibfnamefont {R.}~\bibnamefont
  {Garc{\'\i}a-Muelas}}, \ and\ \bibinfo {author} {\bibfnamefont
  {N.}~\bibnamefont {L{\'o}pez}},\ }\href@noop {} {\bibfield  {journal}
  {\bibinfo  {journal} {Nature Communications}\ }\textbf {\bibinfo {volume}
  {9}},\ \bibinfo {pages} {1} (\bibinfo {year} {2018})}\BibitemShut {NoStop}%
\bibitem [{\citenamefont {Li}\ \emph {et~al.}(2019)\citenamefont {Li},
  \citenamefont {Chiong}, \citenamefont {Hu}, \citenamefont {Cornforth},\ and\
  \citenamefont {Page}}]{Li2019}%
  \BibitemOpen
  \bibfield  {author} {\bibinfo {author} {\bibfnamefont {X.}~\bibnamefont
  {Li}}, \bibinfo {author} {\bibfnamefont {R.}~\bibnamefont {Chiong}}, \bibinfo
  {author} {\bibfnamefont {Z.}~\bibnamefont {Hu}}, \bibinfo {author}
  {\bibfnamefont {D.}~\bibnamefont {Cornforth}}, \ and\ \bibinfo {author}
  {\bibfnamefont {A.~J.}\ \bibnamefont {Page}},\ }\href@noop {} {\bibfield
  {journal} {\bibinfo  {journal} {Journal of Chemical Theory and Computation}\
  }\textbf {\bibinfo {volume} {15}},\ \bibinfo {pages} {6882} (\bibinfo {year}
  {2019})}\BibitemShut {NoStop}%
\bibitem [{\citenamefont {Chang}\ \emph {et~al.}(2019)\citenamefont {Chang},
  \citenamefont {Freeze},\ and\ \citenamefont {Batista}}]{Chang2019}%
  \BibitemOpen
  \bibfield  {author} {\bibinfo {author} {\bibfnamefont {A.~M.}\ \bibnamefont
  {Chang}}, \bibinfo {author} {\bibfnamefont {J.~G.}\ \bibnamefont {Freeze}}, \
  and\ \bibinfo {author} {\bibfnamefont {V.~S.}\ \bibnamefont {Batista}},\
  }\href {\doibase 10.1039/C9SC02339A} {\bibfield  {journal} {\bibinfo
  {journal} {Chemical Science}\ }\textbf {\bibinfo {volume} {10}},\ \bibinfo
  {pages} {6844} (\bibinfo {year} {2019})}\BibitemShut {NoStop}%
\bibitem [{\citenamefont {Beck}(1993)}]{Beck1993}%
  \BibitemOpen
  \bibfield  {author} {\bibinfo {author} {\bibfnamefont {A.~D.}\ \bibnamefont
  {Beck}},\ }\href@noop {} {\bibfield  {journal} {\bibinfo  {journal} {The
  Journal of Chemical Physics}\ }\textbf {\bibinfo {volume} {98}},\ \bibinfo
  {pages} {5648} (\bibinfo {year} {1993})}\BibitemShut {NoStop}%
\bibitem [{\citenamefont {Weigend}\ and\ \citenamefont {Ahlrichs}(2005)}]{SVP}%
  \BibitemOpen
  \bibfield  {author} {\bibinfo {author} {\bibfnamefont {F.}~\bibnamefont
  {Weigend}}\ and\ \bibinfo {author} {\bibfnamefont {R.}~\bibnamefont
  {Ahlrichs}},\ }\href@noop {} {\bibfield  {journal} {\bibinfo  {journal}
  {Physical Chemistry Chemical Physics}\ }\textbf {\bibinfo {volume} {7}},\
  \bibinfo {pages} {3297} (\bibinfo {year} {2005})}\BibitemShut {NoStop}%
\bibitem [{\citenamefont {Weigend}(2006)}]{Weigend2006}%
  \BibitemOpen
  \bibfield  {author} {\bibinfo {author} {\bibfnamefont {F.}~\bibnamefont
  {Weigend}},\ }\href@noop {} {\bibfield  {journal} {\bibinfo  {journal}
  {Physical Chemistry Chemical Physics}\ }\textbf {\bibinfo {volume} {8}},\
  \bibinfo {pages} {1057} (\bibinfo {year} {2006})}\BibitemShut {NoStop}%
\bibitem [{\citenamefont {von Rudorff}\ \emph {et~al.}(2020)\citenamefont {von
  Rudorff}, \citenamefont {Heinen}, \citenamefont {Bragato},\ and\
  \citenamefont {von Lilienfeld}}]{QMRxN}%
  \BibitemOpen
  \bibfield  {author} {\bibinfo {author} {\bibfnamefont {G.~F.}\ \bibnamefont
  {von Rudorff}}, \bibinfo {author} {\bibfnamefont {S.~N.}\ \bibnamefont
  {Heinen}}, \bibinfo {author} {\bibfnamefont {M.}~\bibnamefont {Bragato}}, \
  and\ \bibinfo {author} {\bibfnamefont {O.~A.}\ \bibnamefont {von
  Lilienfeld}},\ }\href@noop {} {\bibfield  {journal} {\bibinfo  {journal}
  {Machine Learning: Science and Technology}\ }\textbf {\bibinfo {volume}
  {1}},\ \bibinfo {pages} {045026} (\bibinfo {year} {2020})}\BibitemShut
  {NoStop}%
\bibitem [{\citenamefont {Meyer}\ \emph {et~al.}(2018)\citenamefont {Meyer},
  \citenamefont {Sawatlon}, \citenamefont {Heinen}, \citenamefont {von
  Lilienfeld},\ and\ \citenamefont {Corminboeuf}}]{Meyer2018}%
  \BibitemOpen
  \bibfield  {author} {\bibinfo {author} {\bibfnamefont {B.}~\bibnamefont
  {Meyer}}, \bibinfo {author} {\bibfnamefont {B.}~\bibnamefont {Sawatlon}},
  \bibinfo {author} {\bibfnamefont {S.}~\bibnamefont {Heinen}}, \bibinfo
  {author} {\bibfnamefont {O.~A.}\ \bibnamefont {von Lilienfeld}}, \ and\
  \bibinfo {author} {\bibfnamefont {C.}~\bibnamefont {Corminboeuf}},\
  }\href@noop {} {\bibfield  {journal} {\bibinfo  {journal} {Chemical Science}\
  }\textbf {\bibinfo {volume} {9}},\ \bibinfo {pages} {7069} (\bibinfo {year}
  {2018})}\BibitemShut {NoStop}%
\bibitem [{\citenamefont {Pizzi}\ \emph {et~al.}(2016)\citenamefont {Pizzi},
  \citenamefont {Cepellotti}, \citenamefont {Sabatini}, \citenamefont
  {Marzari},\ and\ \citenamefont {Kozinsky}}]{aiida}%
  \BibitemOpen
  \bibfield  {author} {\bibinfo {author} {\bibfnamefont {G.}~\bibnamefont
  {Pizzi}}, \bibinfo {author} {\bibfnamefont {A.}~\bibnamefont {Cepellotti}},
  \bibinfo {author} {\bibfnamefont {R.}~\bibnamefont {Sabatini}}, \bibinfo
  {author} {\bibfnamefont {N.}~\bibnamefont {Marzari}}, \ and\ \bibinfo
  {author} {\bibfnamefont {B.}~\bibnamefont {Kozinsky}},\ }\href@noop {}
  {\bibfield  {journal} {\bibinfo  {journal} {Computational Materials Science}\
  }\textbf {\bibinfo {volume} {111}},\ \bibinfo {pages} {218} (\bibinfo {year}
  {2016})}\BibitemShut {NoStop}%
\bibitem [{\citenamefont {Grimme}\ \emph {et~al.}(2010)\citenamefont {Grimme},
  \citenamefont {Antony}, \citenamefont {Ehrlich},\ and\ \citenamefont
  {Krieg}}]{B3LYP2010}%
  \BibitemOpen
  \bibfield  {author} {\bibinfo {author} {\bibfnamefont {S.}~\bibnamefont
  {Grimme}}, \bibinfo {author} {\bibfnamefont {J.}~\bibnamefont {Antony}},
  \bibinfo {author} {\bibfnamefont {S.}~\bibnamefont {Ehrlich}}, \ and\
  \bibinfo {author} {\bibfnamefont {H.}~\bibnamefont {Krieg}},\ }\href@noop {}
  {\bibfield  {journal} {\bibinfo  {journal} {The Journal of Chemical Physics}\
  }\textbf {\bibinfo {volume} {132}},\ \bibinfo {pages} {154104} (\bibinfo
  {year} {2010})}\BibitemShut {NoStop}%
\bibitem [{\citenamefont {Grimme}\ \emph {et~al.}(2011)\citenamefont {Grimme},
  \citenamefont {Ehrlich},\ and\ \citenamefont {Goerigk}}]{B3LYP2011}%
  \BibitemOpen
  \bibfield  {author} {\bibinfo {author} {\bibfnamefont {S.}~\bibnamefont
  {Grimme}}, \bibinfo {author} {\bibfnamefont {S.}~\bibnamefont {Ehrlich}}, \
  and\ \bibinfo {author} {\bibfnamefont {L.}~\bibnamefont {Goerigk}},\
  }\href@noop {} {\bibfield  {journal} {\bibinfo  {journal} {Journal of
  Computational Chemistry}\ }\textbf {\bibinfo {volume} {32}},\ \bibinfo
  {pages} {1456} (\bibinfo {year} {2011})}\BibitemShut {NoStop}%
\bibitem [{\citenamefont {Frisch}\ \emph {et~al.}(2016)\citenamefont {Frisch},
  \citenamefont {Trucks}, \citenamefont {Schlegel}, \citenamefont {Scuseria},
  \citenamefont {Robb}, \citenamefont {Cheeseman}, \citenamefont {Scalmani},
  \citenamefont {Barone}, \citenamefont {Petersson}, \citenamefont {Nakatsuji},
  \citenamefont {Li}, \citenamefont {Caricato}, \citenamefont {Marenich},
  \citenamefont {Bloino}, \citenamefont {Janesko}, \citenamefont {Gomperts},
  \citenamefont {Mennucci}, \citenamefont {Hratchian}, \citenamefont {Ortiz},
  \citenamefont {Izmaylov}, \citenamefont {Sonnenberg}, \citenamefont
  {Williams-Young}, \citenamefont {Ding}, \citenamefont {Lipparini},
  \citenamefont {Egidi}, \citenamefont {Goings}, \citenamefont {Peng},
  \citenamefont {Petrone}, \citenamefont {Henderson}, \citenamefont
  {Ranasinghe}, \citenamefont {Zakrzewski}, \citenamefont {Gao}, \citenamefont
  {Rega}, \citenamefont {Zheng}, \citenamefont {Liang}, \citenamefont {Hada},
  \citenamefont {Ehara}, \citenamefont {Toyota}, \citenamefont {Fukuda},
  \citenamefont {Hasegawa}, \citenamefont {Ishida}, \citenamefont {Nakajima},
  \citenamefont {Honda}, \citenamefont {Kitao}, \citenamefont {Nakai},
  \citenamefont {Vreven}, \citenamefont {Throssell}, \citenamefont
  {Montgomery}, \citenamefont {Peralta}, \citenamefont {Ogliaro}, \citenamefont
  {Bearpark}, \citenamefont {Heyd}, \citenamefont {Brothers}, \citenamefont
  {Kudin}, \citenamefont {Staroverov}, \citenamefont {Keith}, \citenamefont
  {Kobayashi}, \citenamefont {Normand}, \citenamefont {Raghavachari},
  \citenamefont {Rendell}, \citenamefont {Burant}, \citenamefont {Iyengar},
  \citenamefont {Tomasi}, \citenamefont {Cossi}, \citenamefont {Millam},
  \citenamefont {Klene}, \citenamefont {Adamo}, \citenamefont {Cammi},
  \citenamefont {Ochterski}, \citenamefont {Martin}, \citenamefont {Morokuma},
  \citenamefont {Farkas}, \citenamefont {Foresman},\ and\ \citenamefont
  {Fox}}]{g09}%
  \BibitemOpen
  \bibfield  {author} {\bibinfo {author} {\bibfnamefont {M.~J.}\ \bibnamefont
  {Frisch}}, \bibinfo {author} {\bibfnamefont {G.~W.}\ \bibnamefont {Trucks}},
  \bibinfo {author} {\bibfnamefont {H.~B.}\ \bibnamefont {Schlegel}}, \bibinfo
  {author} {\bibfnamefont {G.~E.}\ \bibnamefont {Scuseria}}, \bibinfo {author}
  {\bibfnamefont {M.~A.}\ \bibnamefont {Robb}}, \bibinfo {author}
  {\bibfnamefont {J.~R.}\ \bibnamefont {Cheeseman}}, \bibinfo {author}
  {\bibfnamefont {G.}~\bibnamefont {Scalmani}}, \bibinfo {author}
  {\bibfnamefont {V.}~\bibnamefont {Barone}}, \bibinfo {author} {\bibfnamefont
  {G.~A.}\ \bibnamefont {Petersson}}, \bibinfo {author} {\bibfnamefont
  {H.}~\bibnamefont {Nakatsuji}}, \bibinfo {author} {\bibfnamefont
  {X.}~\bibnamefont {Li}}, \bibinfo {author} {\bibfnamefont {M.}~\bibnamefont
  {Caricato}}, \bibinfo {author} {\bibfnamefont {A.~V.}\ \bibnamefont
  {Marenich}}, \bibinfo {author} {\bibfnamefont {J.}~\bibnamefont {Bloino}},
  \bibinfo {author} {\bibfnamefont {B.~G.}\ \bibnamefont {Janesko}}, \bibinfo
  {author} {\bibfnamefont {R.}~\bibnamefont {Gomperts}}, \bibinfo {author}
  {\bibfnamefont {B.}~\bibnamefont {Mennucci}}, \bibinfo {author}
  {\bibfnamefont {H.~P.}\ \bibnamefont {Hratchian}}, \bibinfo {author}
  {\bibfnamefont {J.~V.}\ \bibnamefont {Ortiz}}, \bibinfo {author}
  {\bibfnamefont {A.~F.}\ \bibnamefont {Izmaylov}}, \bibinfo {author}
  {\bibfnamefont {J.~L.}\ \bibnamefont {Sonnenberg}}, \bibinfo {author}
  {\bibfnamefont {D.}~\bibnamefont {Williams-Young}}, \bibinfo {author}
  {\bibfnamefont {F.}~\bibnamefont {Ding}}, \bibinfo {author} {\bibfnamefont
  {F.}~\bibnamefont {Lipparini}}, \bibinfo {author} {\bibfnamefont
  {F.}~\bibnamefont {Egidi}}, \bibinfo {author} {\bibfnamefont
  {J.}~\bibnamefont {Goings}}, \bibinfo {author} {\bibfnamefont
  {B.}~\bibnamefont {Peng}}, \bibinfo {author} {\bibfnamefont {A.}~\bibnamefont
  {Petrone}}, \bibinfo {author} {\bibfnamefont {T.}~\bibnamefont {Henderson}},
  \bibinfo {author} {\bibfnamefont {D.}~\bibnamefont {Ranasinghe}}, \bibinfo
  {author} {\bibfnamefont {V.~G.}\ \bibnamefont {Zakrzewski}}, \bibinfo
  {author} {\bibfnamefont {J.}~\bibnamefont {Gao}}, \bibinfo {author}
  {\bibfnamefont {N.}~\bibnamefont {Rega}}, \bibinfo {author} {\bibfnamefont
  {G.}~\bibnamefont {Zheng}}, \bibinfo {author} {\bibfnamefont
  {W.}~\bibnamefont {Liang}}, \bibinfo {author} {\bibfnamefont
  {M.}~\bibnamefont {Hada}}, \bibinfo {author} {\bibfnamefont {M.}~\bibnamefont
  {Ehara}}, \bibinfo {author} {\bibfnamefont {K.}~\bibnamefont {Toyota}},
  \bibinfo {author} {\bibfnamefont {R.}~\bibnamefont {Fukuda}}, \bibinfo
  {author} {\bibfnamefont {J.}~\bibnamefont {Hasegawa}}, \bibinfo {author}
  {\bibfnamefont {M.}~\bibnamefont {Ishida}}, \bibinfo {author} {\bibfnamefont
  {T.}~\bibnamefont {Nakajima}}, \bibinfo {author} {\bibfnamefont
  {Y.}~\bibnamefont {Honda}}, \bibinfo {author} {\bibfnamefont
  {O.}~\bibnamefont {Kitao}}, \bibinfo {author} {\bibfnamefont
  {H.}~\bibnamefont {Nakai}}, \bibinfo {author} {\bibfnamefont
  {T.}~\bibnamefont {Vreven}}, \bibinfo {author} {\bibfnamefont
  {K.}~\bibnamefont {Throssell}}, \bibinfo {author} {\bibfnamefont {J.~A.}\
  \bibnamefont {Montgomery}, \bibfnamefont {{Jr.}}}, \bibinfo {author}
  {\bibfnamefont {J.~E.}\ \bibnamefont {Peralta}}, \bibinfo {author}
  {\bibfnamefont {F.}~\bibnamefont {Ogliaro}}, \bibinfo {author} {\bibfnamefont
  {M.~J.}\ \bibnamefont {Bearpark}}, \bibinfo {author} {\bibfnamefont {J.~J.}\
  \bibnamefont {Heyd}}, \bibinfo {author} {\bibfnamefont {E.~N.}\ \bibnamefont
  {Brothers}}, \bibinfo {author} {\bibfnamefont {K.~N.}\ \bibnamefont {Kudin}},
  \bibinfo {author} {\bibfnamefont {V.~N.}\ \bibnamefont {Staroverov}},
  \bibinfo {author} {\bibfnamefont {T.~A.}\ \bibnamefont {Keith}}, \bibinfo
  {author} {\bibfnamefont {R.}~\bibnamefont {Kobayashi}}, \bibinfo {author}
  {\bibfnamefont {J.}~\bibnamefont {Normand}}, \bibinfo {author} {\bibfnamefont
  {K.}~\bibnamefont {Raghavachari}}, \bibinfo {author} {\bibfnamefont {A.~P.}\
  \bibnamefont {Rendell}}, \bibinfo {author} {\bibfnamefont {J.~C.}\
  \bibnamefont {Burant}}, \bibinfo {author} {\bibfnamefont {S.~S.}\
  \bibnamefont {Iyengar}}, \bibinfo {author} {\bibfnamefont {J.}~\bibnamefont
  {Tomasi}}, \bibinfo {author} {\bibfnamefont {M.}~\bibnamefont {Cossi}},
  \bibinfo {author} {\bibfnamefont {J.~M.}\ \bibnamefont {Millam}}, \bibinfo
  {author} {\bibfnamefont {M.}~\bibnamefont {Klene}}, \bibinfo {author}
  {\bibfnamefont {C.}~\bibnamefont {Adamo}}, \bibinfo {author} {\bibfnamefont
  {R.}~\bibnamefont {Cammi}}, \bibinfo {author} {\bibfnamefont {J.~W.}\
  \bibnamefont {Ochterski}}, \bibinfo {author} {\bibfnamefont {R.~L.}\
  \bibnamefont {Martin}}, \bibinfo {author} {\bibfnamefont {K.}~\bibnamefont
  {Morokuma}}, \bibinfo {author} {\bibfnamefont {O.}~\bibnamefont {Farkas}},
  \bibinfo {author} {\bibfnamefont {J.~B.}\ \bibnamefont {Foresman}}, \ and\
  \bibinfo {author} {\bibfnamefont {D.~J.}\ \bibnamefont {Fox}},\ }\href@noop
  {} {\enquote {\bibinfo {title} {Gaussian 09 {R}evision {D}.01},}\ } (\bibinfo
  {year} {2016}),\ \bibinfo {note} {{G}aussian Inc. Wallingford CT}\BibitemShut
  {NoStop}%
\bibitem [{\citenamefont {Rogers}\ and\ \citenamefont {Hahn}(2010)}]{ECFP}%
  \BibitemOpen
  \bibfield  {author} {\bibinfo {author} {\bibfnamefont {D.}~\bibnamefont
  {Rogers}}\ and\ \bibinfo {author} {\bibfnamefont {M.}~\bibnamefont {Hahn}},\
  }\href@noop {} {\bibfield  {journal} {\bibinfo  {journal} {Journal of
  Chemical Information and Modeling}\ }\textbf {\bibinfo {volume} {50}},\
  \bibinfo {pages} {742} (\bibinfo {year} {2010})}\BibitemShut {NoStop}%
\bibitem [{\citenamefont {Hutchinson}\ and\ \citenamefont
  {Kobayashi}(2019)}]{Hutchinson2019}%
  \BibitemOpen
  \bibfield  {author} {\bibinfo {author} {\bibfnamefont {S.~T.}\ \bibnamefont
  {Hutchinson}}\ and\ \bibinfo {author} {\bibfnamefont {R.}~\bibnamefont
  {Kobayashi}},\ }\href@noop {} {\bibfield  {journal} {\bibinfo  {journal}
  {Journal of Chemical Information and Modeling}\ }\textbf {\bibinfo {volume}
  {59}},\ \bibinfo {pages} {1338} (\bibinfo {year} {2019})}\BibitemShut
  {NoStop}%
\bibitem [{\citenamefont {Riniker}(2017)}]{Riniker2017}%
  \BibitemOpen
  \bibfield  {author} {\bibinfo {author} {\bibfnamefont {S.}~\bibnamefont
  {Riniker}},\ }\href@noop {} {\bibfield  {journal} {\bibinfo  {journal}
  {Journal of Chemical Information and Modeling}\ }\textbf {\bibinfo {volume}
  {57}},\ \bibinfo {pages} {726} (\bibinfo {year} {2017})}\BibitemShut
  {NoStop}%
\bibitem [{\citenamefont {Lim}\ and\ \citenamefont {Jung}(2019)}]{Lim2019}%
  \BibitemOpen
  \bibfield  {author} {\bibinfo {author} {\bibfnamefont {H.}~\bibnamefont
  {Lim}}\ and\ \bibinfo {author} {\bibfnamefont {Y.}~\bibnamefont {Jung}},\
  }\href@noop {} {\bibfield  {journal} {\bibinfo  {journal} {Chemical Science}\
  }\textbf {\bibinfo {volume} {10}},\ \bibinfo {pages} {8306} (\bibinfo {year}
  {2019})}\BibitemShut {NoStop}%
\bibitem [{\citenamefont {Lim}\ and\ \citenamefont {Jung}(2021)}]{Lim2021}%
  \BibitemOpen
  \bibfield  {author} {\bibinfo {author} {\bibfnamefont {H.}~\bibnamefont
  {Lim}}\ and\ \bibinfo {author} {\bibfnamefont {Y.}~\bibnamefont {Jung}},\
  }\href@noop {} {\bibfield  {journal} {\bibinfo  {journal} {Journal of
  Cheminformatics}\ }\textbf {\bibinfo {volume} {13}},\ \bibinfo {pages} {1}
  (\bibinfo {year} {2021})}\BibitemShut {NoStop}%
\bibitem [{\citenamefont {Tibshirani}(1996)}]{LASSO}%
  \BibitemOpen
  \bibfield  {author} {\bibinfo {author} {\bibfnamefont {R.}~\bibnamefont
  {Tibshirani}},\ }\href@noop {} {\bibfield  {journal} {\bibinfo  {journal}
  {Journal of the Royal Statistical Society Series B: Statistical Methodology}\
  }\textbf {\bibinfo {volume} {58}},\ \bibinfo {pages} {267} (\bibinfo {year}
  {1996})}\BibitemShut {NoStop}%
\bibitem [{\citenamefont {Chen}\ and\ \citenamefont
  {Guestrin}(2016)}]{xgboost}%
  \BibitemOpen
  \bibfield  {author} {\bibinfo {author} {\bibfnamefont {T.}~\bibnamefont
  {Chen}}\ and\ \bibinfo {author} {\bibfnamefont {C.}~\bibnamefont
  {Guestrin}},\ }in\ \href@noop {} {\emph {\bibinfo {booktitle} {Proceedings of
  the 22nd acm sigkdd international conference on knowledge discovery and data
  mining}}}\ (\bibinfo {year} {2016})\ pp.\ \bibinfo {pages}
  {785--794}\BibitemShut {NoStop}%
\bibitem [{\citenamefont {Abild-Pedersen}\ \emph {et~al.}(2007)\citenamefont
  {Abild-Pedersen}, \citenamefont {Greeley}, \citenamefont {Studt},
  \citenamefont {Rossmeisl}, \citenamefont {Munter}, \citenamefont {Moses},
  \citenamefont {Sk\'ulason}, \citenamefont {Bligaard},\ and\ \citenamefont
  {N\o{}rskov}}]{Abild-Pedersen2007}%
  \BibitemOpen
  \bibfield  {author} {\bibinfo {author} {\bibfnamefont {F.}~\bibnamefont
  {Abild-Pedersen}}, \bibinfo {author} {\bibfnamefont {J.}~\bibnamefont
  {Greeley}}, \bibinfo {author} {\bibfnamefont {F.}~\bibnamefont {Studt}},
  \bibinfo {author} {\bibfnamefont {J.}~\bibnamefont {Rossmeisl}}, \bibinfo
  {author} {\bibfnamefont {T.~R.}\ \bibnamefont {Munter}}, \bibinfo {author}
  {\bibfnamefont {P.~G.}\ \bibnamefont {Moses}}, \bibinfo {author}
  {\bibfnamefont {E.}~\bibnamefont {Sk\'ulason}}, \bibinfo {author}
  {\bibfnamefont {T.}~\bibnamefont {Bligaard}}, \ and\ \bibinfo {author}
  {\bibfnamefont {J.~K.}\ \bibnamefont {N\o{}rskov}},\ }\href {\doibase
  10.1103/PhysRevLett.99.016105} {\bibfield  {journal} {\bibinfo  {journal}
  {Physical Review Letters}\ }\textbf {\bibinfo {volume} {99}},\ \bibinfo
  {pages} {016105} (\bibinfo {year} {2007})}\BibitemShut {NoStop}%
\bibitem [{\citenamefont {Fern{\'{a}}ndez}\ and\ \citenamefont
  {Frenking}(2006)}]{Fernandez2006}%
  \BibitemOpen
  \bibfield  {author} {\bibinfo {author} {\bibfnamefont {I.}~\bibnamefont
  {Fern{\'{a}}ndez}}\ and\ \bibinfo {author} {\bibfnamefont {G.}~\bibnamefont
  {Frenking}},\ }\href {\doibase 10.1021/jo052012e} {\bibfield  {journal}
  {\bibinfo  {journal} {The Journal of Organic Chemistry}\ }\textbf {\bibinfo
  {volume} {71}},\ \bibinfo {pages} {2251} (\bibinfo {year}
  {2006})}\BibitemShut {NoStop}%
\bibitem [{\citenamefont {Calle-Vallejo}\ \emph {et~al.}(2012)\citenamefont
  {Calle-Vallejo}, \citenamefont {Mart{\'\i}nez}, \citenamefont
  {Garc{\'\i}a-Lastra}, \citenamefont {Rossmeisl},\ and\ \citenamefont
  {Koper}}]{Calle2012}%
  \BibitemOpen
  \bibfield  {author} {\bibinfo {author} {\bibfnamefont {F.}~\bibnamefont
  {Calle-Vallejo}}, \bibinfo {author} {\bibfnamefont {J.}~\bibnamefont
  {Mart{\'\i}nez}}, \bibinfo {author} {\bibfnamefont {J.~M.}\ \bibnamefont
  {Garc{\'\i}a-Lastra}}, \bibinfo {author} {\bibfnamefont {J.}~\bibnamefont
  {Rossmeisl}}, \ and\ \bibinfo {author} {\bibfnamefont {M.}~\bibnamefont
  {Koper}},\ }\href@noop {} {\bibfield  {journal} {\bibinfo  {journal}
  {Physical Review Letters}\ }\textbf {\bibinfo {volume} {108}},\ \bibinfo
  {pages} {116103} (\bibinfo {year} {2012})}\BibitemShut {NoStop}%
\bibitem [{\citenamefont {Rupp}\ \emph {et~al.}(2012)\citenamefont {Rupp},
  \citenamefont {Tkatchenko}, \citenamefont {M{\"u}ller},\ and\ \citenamefont
  {von Lilienfeld}}]{rupp2012coulomb}%
  \BibitemOpen
  \bibfield  {author} {\bibinfo {author} {\bibfnamefont {M.}~\bibnamefont
  {Rupp}}, \bibinfo {author} {\bibfnamefont {A.}~\bibnamefont {Tkatchenko}},
  \bibinfo {author} {\bibfnamefont {K.-R.}\ \bibnamefont {M{\"u}ller}}, \ and\
  \bibinfo {author} {\bibfnamefont {O.~A.}\ \bibnamefont {von Lilienfeld}},\
  }\href@noop {} {\bibfield  {journal} {\bibinfo  {journal} {Physical Review
  Letters}\ }\textbf {\bibinfo {volume} {108}},\ \bibinfo {pages} {058301}
  (\bibinfo {year} {2012})}\BibitemShut {NoStop}%
\bibitem [{\citenamefont {Toyao}\ \emph {et~al.}(2018)\citenamefont {Toyao},
  \citenamefont {Suzuki}, \citenamefont {Kikuchi}, \citenamefont {Takakusagi},
  \citenamefont {Shimizu},\ and\ \citenamefont {Takigawa}}]{EP}%
  \BibitemOpen
  \bibfield  {author} {\bibinfo {author} {\bibfnamefont {T.}~\bibnamefont
  {Toyao}}, \bibinfo {author} {\bibfnamefont {K.}~\bibnamefont {Suzuki}},
  \bibinfo {author} {\bibfnamefont {S.}~\bibnamefont {Kikuchi}}, \bibinfo
  {author} {\bibfnamefont {S.}~\bibnamefont {Takakusagi}}, \bibinfo {author}
  {\bibfnamefont {K.-i.}\ \bibnamefont {Shimizu}}, \ and\ \bibinfo {author}
  {\bibfnamefont {I.}~\bibnamefont {Takigawa}},\ }\href@noop {} {\bibfield
  {journal} {\bibinfo  {journal} {The Journal of Physical Chemistry C}\
  }\textbf {\bibinfo {volume} {122}},\ \bibinfo {pages} {8315} (\bibinfo {year}
  {2018})}\BibitemShut {NoStop}%
\bibitem [{\citenamefont {Huang}\ and\ \citenamefont {von
  Lilienfeld}(2020)}]{SLATM}%
  \BibitemOpen
  \bibfield  {author} {\bibinfo {author} {\bibfnamefont {B.}~\bibnamefont
  {Huang}}\ and\ \bibinfo {author} {\bibfnamefont {O.~A.}\ \bibnamefont {von
  Lilienfeld}},\ }\href@noop {} {\bibfield  {journal} {\bibinfo  {journal}
  {Nature Chemistry}\ }\textbf {\bibinfo {volume} {12}},\ \bibinfo {pages}
  {945} (\bibinfo {year} {2020})}\BibitemShut {NoStop}%
\bibitem [{\citenamefont {Hansen}\ \emph {et~al.}(2015)\citenamefont {Hansen},
  \citenamefont {Biegler}, \citenamefont {Ramakrishnan}, \citenamefont
  {Pronobis}, \citenamefont {von Lilienfeld}, \citenamefont {Muller},\ and\
  \citenamefont {Tkatchenko}}]{BoB}%
  \BibitemOpen
  \bibfield  {author} {\bibinfo {author} {\bibfnamefont {K.}~\bibnamefont
  {Hansen}}, \bibinfo {author} {\bibfnamefont {F.}~\bibnamefont {Biegler}},
  \bibinfo {author} {\bibfnamefont {R.}~\bibnamefont {Ramakrishnan}}, \bibinfo
  {author} {\bibfnamefont {W.}~\bibnamefont {Pronobis}}, \bibinfo {author}
  {\bibfnamefont {O.~A.}\ \bibnamefont {von Lilienfeld}}, \bibinfo {author}
  {\bibfnamefont {K.-R.}\ \bibnamefont {Muller}}, \ and\ \bibinfo {author}
  {\bibfnamefont {A.}~\bibnamefont {Tkatchenko}},\ }\href@noop {} {\bibfield
  {journal} {\bibinfo  {journal} {The Journal of Physical Chemistry Letters}\
  }\textbf {\bibinfo {volume} {6}},\ \bibinfo {pages} {2326} (\bibinfo {year}
  {2015})}\BibitemShut {NoStop}%
\bibitem [{\citenamefont {Kubatkin}\ \emph {et~al.}(2003)\citenamefont
  {Kubatkin}, \citenamefont {Danilov}, \citenamefont {Hjort}, \citenamefont
  {Cornil}, \citenamefont {Bredas}, \citenamefont {Stuhr-Hansen}, \citenamefont
  {Hedeg{\aa}rd},\ and\ \citenamefont {Bj{\o}rnholm}}]{Kubatkin2003}%
  \BibitemOpen
  \bibfield  {author} {\bibinfo {author} {\bibfnamefont {S.}~\bibnamefont
  {Kubatkin}}, \bibinfo {author} {\bibfnamefont {A.}~\bibnamefont {Danilov}},
  \bibinfo {author} {\bibfnamefont {M.}~\bibnamefont {Hjort}}, \bibinfo
  {author} {\bibfnamefont {J.}~\bibnamefont {Cornil}}, \bibinfo {author}
  {\bibfnamefont {J.-L.}\ \bibnamefont {Bredas}}, \bibinfo {author}
  {\bibfnamefont {N.}~\bibnamefont {Stuhr-Hansen}}, \bibinfo {author}
  {\bibfnamefont {P.}~\bibnamefont {Hedeg{\aa}rd}}, \ and\ \bibinfo {author}
  {\bibfnamefont {T.}~\bibnamefont {Bj{\o}rnholm}},\ }\href@noop {} {\bibfield
  {journal} {\bibinfo  {journal} {Nature}\ }\textbf {\bibinfo {volume} {425}},\
  \bibinfo {pages} {698} (\bibinfo {year} {2003})}\BibitemShut {NoStop}%
\bibitem [{\citenamefont {Roncali}(2007)}]{Roncali2007}%
  \BibitemOpen
  \bibfield  {author} {\bibinfo {author} {\bibfnamefont {J.}~\bibnamefont
  {Roncali}},\ }\href@noop {} {\bibfield  {journal} {\bibinfo  {journal}
  {Macromolecular Rapid Communications}\ }\textbf {\bibinfo {volume} {28}},\
  \bibinfo {pages} {1761} (\bibinfo {year} {2007})}\BibitemShut {NoStop}%
\bibitem [{\citenamefont {Jurow}\ \emph {et~al.}(2010)\citenamefont {Jurow},
  \citenamefont {Schuckman}, \citenamefont {Batteas},\ and\ \citenamefont
  {Drain}}]{Jurow2010}%
  \BibitemOpen
  \bibfield  {author} {\bibinfo {author} {\bibfnamefont {M.}~\bibnamefont
  {Jurow}}, \bibinfo {author} {\bibfnamefont {A.~E.}\ \bibnamefont
  {Schuckman}}, \bibinfo {author} {\bibfnamefont {J.~D.}\ \bibnamefont
  {Batteas}}, \ and\ \bibinfo {author} {\bibfnamefont {C.~M.}\ \bibnamefont
  {Drain}},\ }\href@noop {} {\bibfield  {journal} {\bibinfo  {journal}
  {Coordination Chemistry Reviews}\ }\textbf {\bibinfo {volume} {254}},\
  \bibinfo {pages} {2297} (\bibinfo {year} {2010})}\BibitemShut {NoStop}%
\bibitem [{\citenamefont {Tao}\ \emph {et~al.}(2017)\citenamefont {Tao},
  \citenamefont {Cao},\ and\ \citenamefont {Bobbert}}]{Tao2017}%
  \BibitemOpen
  \bibfield  {author} {\bibinfo {author} {\bibfnamefont {S.~X.}\ \bibnamefont
  {Tao}}, \bibinfo {author} {\bibfnamefont {X.}~\bibnamefont {Cao}}, \ and\
  \bibinfo {author} {\bibfnamefont {P.~A.}\ \bibnamefont {Bobbert}},\
  }\href@noop {} {\bibfield  {journal} {\bibinfo  {journal} {Scientific
  Reports}\ }\textbf {\bibinfo {volume} {7}},\ \bibinfo {pages} {1} (\bibinfo
  {year} {2017})}\BibitemShut {NoStop}%
\bibitem [{\citenamefont {Stoliaroff}\ and\ \citenamefont
  {Latouche}(2020)}]{Stoliaroff2020}%
  \BibitemOpen
  \bibfield  {author} {\bibinfo {author} {\bibfnamefont {A.}~\bibnamefont
  {Stoliaroff}}\ and\ \bibinfo {author} {\bibfnamefont {C.}~\bibnamefont
  {Latouche}},\ }\href@noop {} {\bibfield  {journal} {\bibinfo  {journal} {The
  Journal of Physical Chemistry C}\ }\textbf {\bibinfo {volume} {124}},\
  \bibinfo {pages} {8467} (\bibinfo {year} {2020})}\BibitemShut {NoStop}%
\bibitem [{\citenamefont {Mazouin}\ \emph {et~al.}(2022)\citenamefont
  {Mazouin}, \citenamefont {Sch{\"o}pfer},\ and\ \citenamefont {von
  Lilienfeld}}]{mazouin2022selected}%
  \BibitemOpen
  \bibfield  {author} {\bibinfo {author} {\bibfnamefont {B.}~\bibnamefont
  {Mazouin}}, \bibinfo {author} {\bibfnamefont {A.~A.}\ \bibnamefont
  {Sch{\"o}pfer}}, \ and\ \bibinfo {author} {\bibfnamefont {O.~A.}\
  \bibnamefont {von Lilienfeld}},\ }\href@noop {} {\bibfield  {journal}
  {\bibinfo  {journal} {Materials Advances}\ }\textbf {\bibinfo {volume} {3}},\
  \bibinfo {pages} {8306} (\bibinfo {year} {2022})}\BibitemShut {NoStop}%
\bibitem [{\citenamefont {Yukawa}\ and\ \citenamefont
  {Tsuno}(1959)}]{yukawa1959}%
  \BibitemOpen
  \bibfield  {author} {\bibinfo {author} {\bibfnamefont {Y.}~\bibnamefont
  {Yukawa}}\ and\ \bibinfo {author} {\bibfnamefont {Y.}~\bibnamefont {Tsuno}},\
  }\href {\doibase 10.1246/bcsj.32.965} {\bibfield  {journal} {\bibinfo
  {journal} {Bulletin of the Chemical Society of Japan}\ }\textbf {\bibinfo
  {volume} {32}},\ \bibinfo {pages} {965} (\bibinfo {year} {1959})}\BibitemShut
  {NoStop}%
\end{thebibliography}%

\end{document}